\documentclass[twocolumn,showpacs,preprintnumbers,superscriptaddress,amsfonts,prd,floatfix,tightenlines]{revtex4}
\usepackage{graphicx}
\usepackage{dcolumn}
\usepackage{longtable}%
\usepackage[dvips]{epsfig}
\usepackage{bm}

\newcommand{\ycm}{\mbox{$y_{\rm{cm}}$}}

\def\PRETH{{\sc pretrigger hi}}
\def\GLL{{\sc local$\otimes$global lo}}
\def\SLL{{\sc single local lo}}
\def\SLH{{\sc single local hi}}

\def\LOCALH{{\sc local hi}}
\def\LOCALL{{\sc local lo}}
\def\GLOBALL{{\sc global lo}}

\def\gb{$\gamma_{bckg}$}

\def\pBe{${\mit p}$Be}
\def\pp{${\mit pp}$}
\def\pip{$\pi^{-}\mit{p}$}
\def\piBe{${\pi^{-}}$Be}

\def\DIFFXS{Ed \sigma/d^{3}p}

\def\UCDAVIS{University of California-Davis, Davis, California 95616}
\def\MSU{Michigan State University, East Lansing, Michigan 48824}
\def\Delhi{University of Delhi, Delhi, India 110007}
\def\FNAL{Fermi National Accelerator Laboratory, Batavia,
                   Illinois 60510}
\def\NEU{Northeastern University, Boston, Massachusetts  02115}
\def\OK{University of Oklahoma, Norman, Oklahoma  73019}
\def\PSU{Pennsylvania State University, University Park,
		   Pennsylvania 16802}
\def\PU{University of Pittsburgh, Pittsburgh, Pennsylvania 15260}
\def\UR{University of Rochester, Rochester, New York 14627}

\def\bbcoord{20 150 652 652} 
\def\bbcoordb{20 400 652 652} 
\def\bbcoordd{50 400 652 675}
\def\figsize{4.2in}

\begin{document}

\preprint{FERMILAB-Pub-04/035-E}
\title{
Measurement of direct photon production at Tevatron fixed target
energies}
\author{L.~Apanasevich}\affiliation{\MSU}\affiliation{\UR}
\author{J.~Bacigalupi}\affiliation{\UCDAVIS}
\author{W.~Baker}\affiliation{\FNAL}
\author{M.~Begel}\affiliation{\UR}
\author{S.~Blusk}\affiliation{\PU}
\author{C.~Bromberg}\affiliation{\MSU}
\author{P.~Chang}\affiliation{\NEU}
\author{B.~Choudhary}\affiliation{\Delhi}
\author{W.~H.~Chung}\affiliation{\PU}
\author{L.~de~Barbaro}\affiliation{\UR}
\author{W.~DeSoi}\affiliation{\UR}
\author{W.~D{\l}ugosz}\affiliation{\NEU}
\author{J.~Dunlea}\affiliation{\UR}
\author{E.~Engels,~Jr.}\affiliation{\PU}
\author{G.~Fanourakis}\affiliation{\UR}
\author{T.~Ferbel}\affiliation{\UR}
\author{J.~Ftacnik}\affiliation{\UR}
\author{D.~Garelick}\affiliation{\NEU}
\author{G.~Ginther}\affiliation{\UR}
\author{M.~Glaubman}\affiliation{\NEU}
\author{P.~Gutierrez}\affiliation{\OK}
\author{K.~Hartman}\affiliation{\PSU}
\author{J.~Huston}\affiliation{\MSU}
\author{C.~Johnstone}\affiliation{\FNAL}
\author{V.~Kapoor}\affiliation{\Delhi}
\author{J.~Kuehler}\affiliation{\OK}
\author{C.~Lirakis}\affiliation{\NEU}
\author{F.~Lobkowicz}\altaffiliation{Deceased}\affiliation{\UR}
\author{P.~Lukens}\affiliation{\FNAL}
\author{J.~Mansour}\affiliation{\UR}
\author{A.~Maul}\affiliation{\MSU}
\author{R.~Miller}\affiliation{\MSU}
\author{B.~Y.~Oh}\affiliation{\PSU}
\author{G.~Osborne}\affiliation{\UR}
\author{D.~Pellett}\affiliation{\UCDAVIS}
\author{E.~Prebys}\affiliation{\UR}
\author{R.~Roser}\affiliation{\UR}
\author{P.~Shepard}\affiliation{\PU}
\author{R.~Shivpuri}\affiliation{\Delhi}
\author{D.~Skow}\affiliation{\FNAL}
\author{P.~Slattery}\affiliation{\UR}
\author{L.~Sorrell}\affiliation{\MSU}
\author{D.~Striley}\affiliation{\NEU}
\author{W.~Toothacker}\altaffiliation{Deceased}\affiliation{\PSU}
\author{S.~M.~Tripathi}\affiliation{\UCDAVIS}
\author{N.~Varelas}\affiliation{\UR}
\author{D.~Weerasundara}\affiliation{\PU}
\author{J.~J.~Whitmore}\affiliation{\PSU}
\author{T.~Yasuda}\affiliation{\NEU}
\author{C.~Yosef}\affiliation{\MSU}
\author{M.~Zieli\'{n}ski}\affiliation{\UR}
\author{V.~Zutshi}\affiliation{\Delhi}
\collaboration{Fermilab E706 Collaboration}\noaffiliation

\date{\today}

\begin{abstract}
Measurements of the production of high transverse momentum
direct photons by a 515~GeV/$c$ $\pi^-$ beam and 530 and 800~GeV/$c$
proton beams in interactions with beryllium and hydrogen targets
are presented.  The
data span the kinematic ranges of $3.5 < p_T < 12$~GeV/$c$ in
transverse momentum and 1.5 units in rapidity.  The inclusive
direct-photon cross sections are compared with next-to-leading-order
perturbative QCD calculations 
and expectations based on a
phenomenological parton-$k_T$ model.
\end{abstract}
\pacs{13.85.Qk, 12.38.Qk}

\maketitle

\section{Introduction}

Inclusive single-particle production at large transverse
\begin{figure}
\epsfxsize=\figsize
\epsfbox[100 350 625 500]{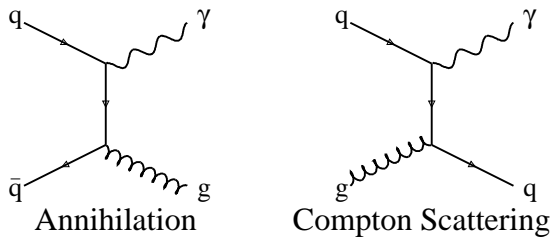}
\caption{Leading-order diagrams for direct-photon production.}
\label{fig:feynman}
\end{figure}
momentum ($p_T$) has been useful in the development of perturbative
quantum chromodynamics (PQCD)~\cite{geist,mccubbin,owens,molzon}.  
Quantitative comparisons with
data have yielded information on the validity of the PQCD
description, and on parton distribution functions of hadrons (PDF)
and the fragmentation functions of partons.  This paper reports
precision measurements of the production of direct photons with
large $p_T$.

At leading order, only two processes contribute to the direct-photon
cross section, namely, $q\bar{q}$ annihilation and quark--gluon
Compton scattering (Fig.~\ref{fig:feynman}).  The photon's momentum
reflects the collision kinematics since such photons are produced at
the elementary interaction vertex.  This contrasts with jet production,
where the hadronization process obscures the measurement of energy
and direction of the outgoing parton.  A complete theoretical
description of the direct-photon process is of special importance as
it has long been expected to facilitate the extraction of
the gluon distribution of the proton.
The quark--gluon Compton
scattering process shown in Fig.~\ref{fig:feynman} provides a major
contribution to inclusive direct-photon production.  The gluon
distribution is relatively well constrained by deep-inelastic and
Drell-Yan data
for momentum fractions
$x<0.1$, but less so at
larger~$x$~\cite{huston-uncertainty}.  Fixed-target direct-photon data
were incorporated into several previous global parton-distribution
analyses~\cite{cteq4,grv92,mrst98} to constrain the gluon distribution
at large~$x$, but more recent global PDF analyses have not included such
photon data; this point is revisited in the concluding section of
this work.

\section{Apparatus}

\subsection{Meson West spectrometer}

Fermilab E706 is a fixed-target experiment designed to measure the
production of direct photons, neutral mesons, and associated particles
at
high-$p_T$~\cite{E706-kt,E706-pos-pieta,E706-neg-pieta,E706-charm,E706-omega}.
The apparatus included a charged particle spectrometer, a large liquid
argon calorimeter and downstream muon identifiers. 

The data collection phase of this experiment spanned three fixed-target
running periods, including the relatively low-statistics commissioning
run during 1987-1988~\cite{E706-88,E706-88-recoil}, 
and the two primary data runs 
in 1990 and 1991-1992.
This paper reports on data from the two primary data runs of the
experiment~\cite{note-88}.  During the 1990 run, the target
consisted of two 0.8~mm thick copper foils followed by two pieces of
beryllium.  The upstream piece was 3.7~cm long, while the length of
the downstream piece was 1.1~cm.  In the 1991-1992 run, the target
consisted of two 0.8~mm thick copper foils immediately upstream of a
liquid hydrogen target~\cite{h2target}, followed by a 2.54~cm long
beryllium cylinder.  The liquid hydrogen was contained in a 15.3~cm
long mylar flask, which was supported in an evacuated volume with
beryllium windows at each end (2.5~mm thickness upstream and 2.8~mm
thickness downstream).

The charged particle spectrometer consisted of
silicon microstrip detectors (SSDs) in the target region and multiwire
proportional chambers (PWCs) and straw tube drift chambers (STDCs)
downstream of a large-aperture analysis magnet~\cite{E706-charm}.
Six 3$\times$3~cm$^2$ SSD planes were located upstream of the target region
and used to reconstruct beam tracks.  Two hybrid 5$\times $5~cm$^2$
SSD planes (25~$\mu$m pitch strips in the central 1~cm, 50~$\mu$m
beyond) were located downstream of the target region. These were followed by
eight 5$\times $5~cm$^2$ SSD planes of 50~$\mu$m pitch.  The analysis
dipole imparted a $0.45~{\rm GeV}/c$ $p_T$ impulse in the
horizontal plane to charged particles.  Downstream track segments were
measured by means of four stations of four views ($XYUV$) of 2.54~mm
pitch PWCs and two stations of eight (4$X$4$Y$) layers of STDCs with
tube diameters 1.03~cm (upstream station) and 1.59~cm (downstream
station)~\cite{E706-STDC}.

\begin{figure}
\epsfxsize=4in
\vskip1.3cm
\epsfbox[0 72 612 720]{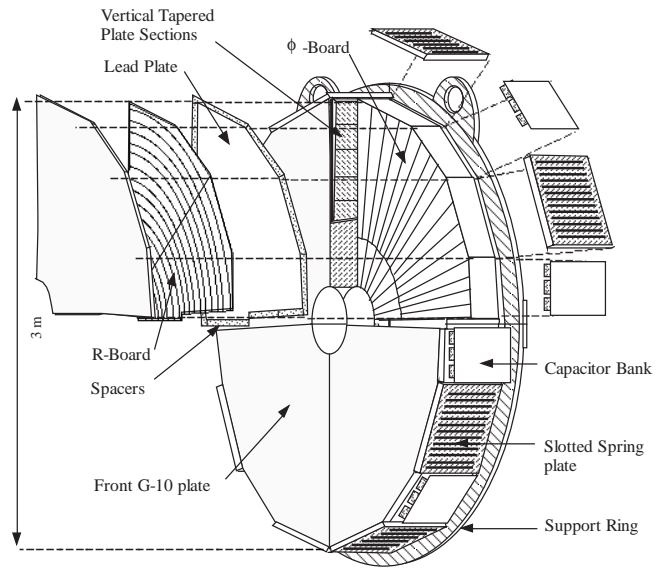}
\vskip-4.0cm
\caption{
A drawing of the 
liquid argon electromagnetic calorimeter with some components
pulled away in one quadrant to reveal a view of the internal details.}
\label{fig:emlac}
\end{figure}

Photons were detected in a large, lead and liquid-argon sampling
electromagnetic calorimeter (EMLAC), located 9~m downstream of the
target~\cite{E706-calibration}.  The EMLAC had a cylindrical geometry
with an inner radius of 20~cm and an outer radius of 160~cm.  The
calorimeter had 33~longitudinal cells read out in two sections: an 11
cell front section (8.5~radiation lengths) and a 22 cell 
back section (18~radiation lengths).  Each longitudinal cell 
consisted of a 2~mm thick lead cathode (the first cathode was
constructed of aluminum), a double-sided copper-clad G-10 radial ($R$) anode
board, a second 2~mm thick lead cathode, and a double-sided
copper-clad G-10 azimuthal ($\Phi$) anode board.  
The 2.5~mm gaps between these layers were filled with liquid argon.
The physical layout is illustrated in
Fig.~\ref{fig:emlac}.

The EMLAC readout was subdivided azimuthally into octants, each
consisting of interleaved, finely segmented, radial and 
azimuthal 
views.  This segmentation was realized by cutting the
copper-cladding on the anode boards to form either radial or azimuthal
strips.  Signals from corresponding strips from all $R$ (or $\Phi$)
anode boards in the front (or back) section of a given octant
were jumpered together.  The copper-cladding on the  radial anode
boards was cut into concentric strips centered on the nominal beam
axis.  The width of the strips on the first $R$ board was 5.5~mm.
The width of the strips on the following $R$ boards increased 
slightly so that the radial geometry was projective relative to the
target region.
The azimuthal strips were
split at a radius of 40~cm into inner and outer segments;
each inner strip subtended an azimuthal angle of $\pi/192$~radians,
while outer strips covered $\pi/384$~radians.

The spectrometer also included two other calorimeters: a hadronic
calorimeter located downstream of the EMLAC within the same cryostat,
and a steel and scintillator calorimeter positioned further downstream
to increase coverage in the very forward region.  The E672 muon
spectrometer, consisting of a toroidal magnet, shielding, scintillator, and
proportional wire chambers, was deployed immediately downstream of the
calorimeters~\cite{psi90,psi91,chi90,B90}.

The spectrometer was located at the end of the Meson West beamline.
The design of the beamline, primary target and 
primary beam dump were intended to minimize the rate of 
beam-halo muons incident upon the spectrometer.
The beamline was capable of transporting either a 
primary (800~GeV/$c$) proton beam or
unseparated secondary particle beams of either polarity
to the experimental hall.  The beamline \v{C}erenkov detector
was used to 
identify the secondary beam particles~\cite{striley}.  This
43.4~m long helium-filled counter was located 100~m upstream
of the experimental target. 
The positive secondary beam with mean momentum of 530~GeV/$c$ was
97\% protons.
The negative secondary beam with mean momentum of 515~GeV/$c$ was
99\% pions.

At the end of the beamline was a 4.7~m
long stack of steel surrounding the beam pipe and shadowing the EMLAC
to absorb off-axis hadrons.  A water tank was placed at the downstream
end of this hadron shield to absorb low-energy neutrons. Surrounding
the hadron shield and neutron absorber were walls of scintillation
counters (VW) to identify penetrating muons.  There was one wall at
the upstream end and two walls at the downstream end of the hadron
absorber during the 1990 run. An additional wall was added to the
upstream end of the hadron absorber prior to the 1991-1992 run.

\subsection{Trigger}

The E706 trigger selected interactions yielding high-$p_T$ showers in the
EMLAC. The selection process involved several stages: beam and
interaction definitions, a pretrigger, and high-$p_T$ trigger
requirements~\cite{E706-trigger,E706-charm,E706-pos-pieta}.  A
scintillator hodoscope, located 2~m upstream of the target
region, was used to detect beam particles, and reject interactions
with more than one incident beam particle.  
Additional scintillator with a 1~cm diameter central hole 
was located just downstream of the
beam hodoscope, 
and served to reject
interactions initiated by particles in the beam halo~\cite{BH}.  
Two pairs of scintillator
counters, mounted on the dipole analysis magnet, were used to identify
interactions in the target.  
A filter was
employed for the high-$p_T$ triggers to reject interactions that 
occurred within
60~ns of one another to minimize potential confusion in the EMLAC due
to out-of-time interactions.

For those interactions that satisfied the beam and interaction
requirements, the $p_T$ deposited in various regions of the EMLAC
was evaluated by weighting the 
energy signals from the EMLAC $R$-channel amplifier fast outputs
by a factor proportional to $\sin{\theta_i}$, where
$\theta_i$ was the polar angle between the $i^{th}$ strip
and the nominal beam axis.  The
\PRETH\ requirement for a given octant  was satisfied when the
$p_T$ detected in
either the inner 128 $R$ channels 
or the outer $R$ channels of that octant was greater than
a threshold value.  A pretrigger signal was issued only when there was no
evidence in that octant of substantial noise, significant $p_T$
attributable to an earlier interaction, or incident beam-halo muon
detected by the VW.

Localized trigger groups were formed for each octant by clustering the
$R$-channel fast-outputs into 32~groups of 8~channels.  Each adjacent
pair of 8~channel groups formed a group-of-16 strips.  If the $p_T$
detected in any of these groups-of-16 was above a specified high (or
low) threshold, then a \LOCALH\ (or \LOCALL) signal was generated for
that octant.  A \SLH\ (or \SLL) trigger was generated if a \LOCALH\
(or \LOCALL) signal was generated in coincidence with the \PRETH\ in
the same octant.  

Trigger decisions were also made based upon global energy depositions
within an octant.  A \GLOBALL\ signal was generated if the total $p_T$ in
an octant exceeded a threshold value.  The \GLL\ trigger required a
coincidence of the \PRETH\ signal with \GLOBALL\ and
\LOCALL\ signals from the same octant. The \LOCALL\ requirement was
included to suppress spurious global triggers due to coherent noise in
the EMLAC.

The \SLL\ and \GLL\ triggers were prescaled to keep them from
dominating the data sample. Prescaled samples of beam, interaction,
and pretrigger events were also recorded.

\section{Analysis Methods}

Data samples contributing to this analysis represent an integrated
luminosity of~6.8 (1.1)~pb$^{-1}$ for 530~GeV/$c$ $p$Be ($pp$)
interactions, and~6.5 (1.1)~pb$^{-1}$ at 800~GeV/$c$.  These samples
were accumulated during the 1991-1992 run.  
Results reported in this paper also use 
6.1~pb$^{-1}$ of $\pi^-$Be data recorded during the 1990 run,  as well as 
1.4~pb$^{-1}$ of $\pi^-$Be data 
and 0.23~pb$^{-1}$ of $\pi^-p$ data accumulated during the 1991-1992 run.
The following subsections
describe the analysis procedures and methods used to correct the
data for losses due to inefficiencies and selection biases.
Additional details can be found in several of our previous
papers~\cite{E706-pos-pieta,E706-neg-pieta,E706-charm,E706-calibration}.

\subsection{Event reconstruction}

Two major aspects of the analysis procedure involved charged-particle
and calorimeter shower reconstruction.  The charged-track
reconstruction algorithm produced track segments upstream of the
magnet using information from the SSDs, and downstream of the magnet
using information from the PWCs and STDCs. These track segments
were projected to the center of the magnet and linked to form the
final tracks whose calculated charges and momenta were used for the
physics analysis.  The charged track reconstruction is described in
more detail elsewhere~\cite{E706-charm}.

The readout in each EMLAC quadrant consisted of four regions: left and
right $R$, and inner and outer~$\Phi$.  Strip energies from clusters
in each region were fit to the shape of an electromagnetic shower
determined from detailed Monte Carlo simulations and isolated-shower data.
These fits were used to evaluate the positions and energies of the
peaks in each region.  Shower positions and energies were obtained by
correlating peaks of approximately the same energy in the $R$ and
$\Phi$ regions within the same half octant.  More complex algorithms
were used to handle configurations with showers spanning multiple
regions.  The EMLAC readout was also subdivided longitudinally into
front and back sections.  This segmentation provided discrimination
between showers generated by electromagnetically or hadronically
interacting particles.  An expanded discussion of the EMLAC
reconstruction procedures and performance can be found 
elsewhere~\cite{E706-calibration}.

\begin{figure}[t]
\epsfxsize=\figsize
\epsfbox[\bbcoord]{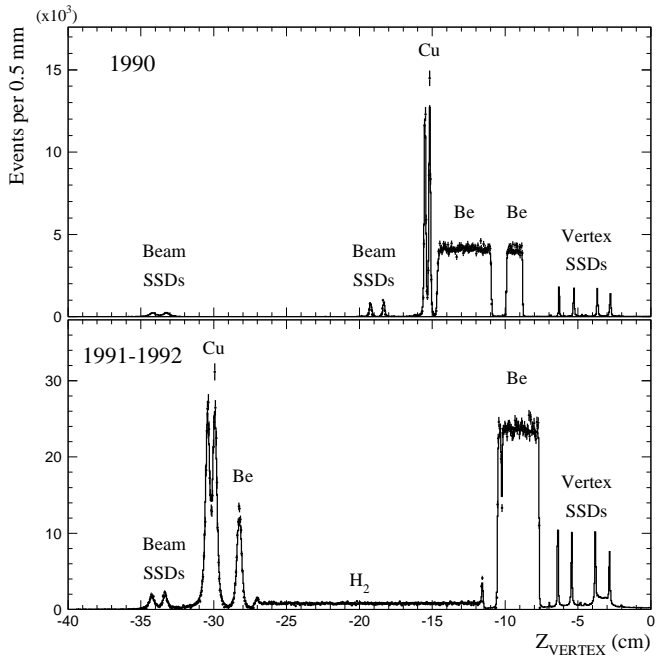}
\caption{The longitudinal distribution of reconstructed vertices for 
1990~(top) and 1991-1992~(bottom) target configurations.  The data are
represented by the histogram, the results from the detailed
detector simulation by the points.}
\label{fig:vz}
\end{figure}

\subsection{Selection criteria and corrections}

Only events with reconstructed vertices within the
fiducial volume of the Be or H${}_2$ targets
contribute to the results presented in this paper.  Vertex reconstruction
efficiencies were evaluated for each target using a detailed Monte
Carlo simulation of the spectrometer, 
as described in Section~III.F below~\cite{herwig61}.
These
efficiencies were used to correct for reconstruction losses and resolution
smearing across fiducial boundaries of the targets.  
The corrections were $<$~1\% for interactions in
the beryllium targets in the 1990 configuration
as well as for the hydrogen and downstream beryllium targets in the
1991-1992 configuration.
The correction was $<$3.5\% for the upstream Be target
in the 1991-1992 configuration. 
The longitudinal distribution of reconstructed vertices are shown, for
both 1990 and 1991-1992 target configurations, in Fig.~\ref{fig:vz}.
Overlaid on the data for the copper, beryllium, and hydrogen targets
are the results from the detailed Monte Carlo simulation.

Only photons that were reconstructed with the fiducial regions of
the EMLAC contributed to the cross-section results presented.
In particular, regions of the detector near quadrant boundaries that
abutted steel support plates, the central beam hole, the outer radius
of the EMLAC, and octant boundaries were excluded.  A simple
ray-tracing Monte Carlo program was employed to determine the
correction for the $\Phi$-dependent fiducial requirements.

\begin{figure}
\epsfxsize=\figsize
\epsfbox[\bbcoord]{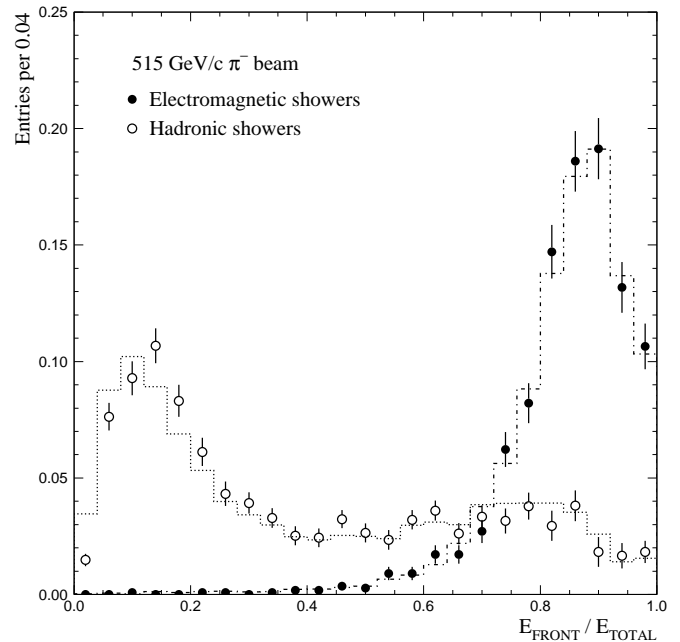}
\caption{Fraction of energy deposited in the front section of the EMLAC 
for identified electromagnetic showers (electrons from photon
conversions) and hadronic showers (charged pions from $K_s$ decays).
The data are represented by the histograms, the results from the
detailed detector simulation by the points. The distributions have
been normalized to unit area.}
\label{fig:efet}
\end{figure}

Showers that deposited  
$\leq$20\% of their energy in the front section of the EMLAC
were not considered as photon candidates since such
showers were likely to be due to hadronic
interactions~\cite{E706-calibration} (Fig.~\ref{fig:efet}). 
Losses of photons
due to this requirement were $\approx2$\%.  A detailed event 
simulation (described in Section~III.F below) 
was employed to correct for this and
other effects including reconstruction smearing and losses.

\begin{figure}[t]
\epsfxsize=\figsize
\epsfbox[\bbcoordd]{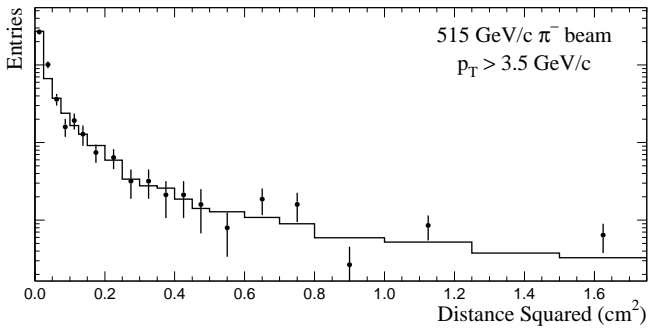}
\caption{Distribution of the distance squared between the
reconstructed positions of electromagnetic
showers and the nearest charged-particle 
track in data (histogram) and in the 
detailed detector
simulation ($\bullet$). 
To enhance the electron fraction in these samples,
only showers with $>0.75$ of their energy deposited
in the front section of the EMLAC contribute to this plot.
The distributions have been normalized to the same 
area.}
\label{fig:dtrk}
\end{figure}

Reconstructed showers were excluded from the photon 
sample when 
charged-particle tracks pointed to within 1~cm of shower center
(Fig.~\ref{fig:dtrk}). 
The correction for this criterion in the direct-photon analysis is
$\approx1.01$ based upon studies of the impact of this requirement on
reconstructed $\pi^0$'s.

The correction for losses due to the conversion of photons into
$e^+e^-$ pairs was evaluated by projecting each reconstructed photon
from the event vertex to the reconstructed position in the EMLAC.  The
radiation length of material traversed, up to the analysis magnet, was
evaluated based upon detailed detector descriptions.  The photon
conversion probability was evaluated and used to account for
conversion losses. The average correction for conversion losses was
$1.09$ per photon for the Be target in the 1990 run 
($1.08$ in 1991-1992 run) and
$1.11$ per photon for the H${}_2$ target.

\subsection{Trigger corrections}

Trigger corrections were evaluated on an event-by-event basis
using the energy depositions detected in the EMLAC combined 
with the measured responses of the trigger groups in which those
energy depositions occurred for this particular trigger and event. 
The probability~($p_{\mit i}$)
that the ${\mit i}^{th}$ trigger group (e.g.,
a group-of-16 or inner 128~$R$ channels) in the octant was satisfied
was measured as a function of the $p_T$ reconstructed within the group
using data samples unbiased with respect to that
group~\cite{E706-trigger}.  The trigger probability for a given octant
was defined as $P = 1 -\prod(1-p_{\mit i})$.
Trigger corrections were evaluated on an event-by-event basis by
applying the inverse of the trigger probability as a weight to each
photon candidate. Photon candidates with $P < 0.1$ were excluded from
further consideration to avoid excessively large weights.  The
correction for losses from this requirement was determined from the detailed
detector simulation and absorbed into the reconstruction efficiency
described below.

The results presented in this paper are extracted from 
data selected using several triggers with different
thresholds and prescale factors 
to probe a broad range of $p_T$.
The lower threshold and prescaled triggers are used to populate
the lower $p_T$ range where the data from the 
higher threshold triggers were judged to be less reliable 
(due to larger average trigger corrections) than corresponding
data from the lower
threshold triggers.  These judgements were made by 
comparing the fully corrected results from each trigger in $p_T$ and
rapidity, \ycm.  These transitions were functions of both
$p_T$ and \ycm, and were different for the three samples.  The \SLH\
trigger was used exclusively for photons with $p_T>4.0$~GeV/$c$.

\subsection{Rejection of beam-halo muons}

Spurious triggers were produced by muons in the beam halo that
deposited significant energy in the EMLAC in random coincidence with 
interactions in the target.  Such muon-induced showers
were interpreted by the trigger as high-$p_T$ depositions,
particularly when they occurred in the outer regions of the EMLAC.
While the pretrigger logic
used VW signals to reject events associated with such muons,
additional requirements on the latched VW signals, the total $p_T$
imbalance in the event, and the direction and shape of reconstructed
showers were imposed during the offline analysis 
to further suppress this background.
We note that these criteria successfully eliminated the beam-halo 
muon background in the high-$p_T$
meson samples  
(see Fig.~6 of Ref.~\cite{E706-pos-pieta}
and Fig.~5 of Ref.~\cite{E706-neg-pieta}).  As expected,
the beam-halo muons rates were much larger in the
secondary beam data samples than in the primary beam (800~GeV/$c$) data sample.

The $p_T$ balance in the event was defined by calculating the $p_T$ of
the photons and charged particles which, based upon their initial
trajectories, would intercept the EMLAC in the transverse plane within
the $120^\circ$ sector opposite the direct-photon candidate.  The
ratio of this ``away-side'' $p_T$ ($P_T^{away}$) to that of the 
direct-photon candidate
should be near unity in interactions that generate a high-$p_T$
photon. However, for events triggered by showers from beam-halo muons,
$P_T^{away}/p_T$ should be near zero, since interactions 
are typically soft. 
Showers with $P_T^{away}/p_T < 0.3$ were 
excluded from the photon sample.

A measure of the incident direction of showers observed in the EMLAC
was characterized as $\delta =
{R_\mathrm{front}-\left(Z_\mathrm{front}/Z_\mathrm{back}\right)
R_\mathrm{back}}$, where ${R_\mathrm{front}}$ and ${R_\mathrm{back}}$
are the radial positions of the reconstructed shower in the front and
back sections, and ${Z_\mathrm{front}}$ and ${Z_\mathrm{back}}$ are
the longitudinal locations of those EMLAC sections relative to the
nominal target location.  Showers with
$\delta\sim0$ were generated by particles from the target region, while
showers with large $\delta$ were generated by particles incident from
other directions.  The requirement on $\delta$ was a function of
radius; showers with $\delta>0.193$~cm for $R<40$~cm ($\delta>0.0048
R$ for $R\geq 40$~cm) were excluded from the photon 
sample.

The electromagnetic shower shape 
employed in the photon reconstruction program was determined for photons
originating in the target region~\cite{E706-calibration}.
Consequently, showers resulting from beam-halo muons were frequently poorly fit
by this shower shape, particularly in the radial view. The $\chi^2$ of
the radial view fit therefore provided additional discrimination against 
beam-halo muons.  Showers with $\chi^2_R/E > 0.1$~GeV$^{-1}$ 
(where $E$ represents the
shower energy) were excluded from the
photon sample.

\begin{figure}
\epsfxsize=\figsize
\vskip0.15truein
\epsfbox[60 150 652 652]{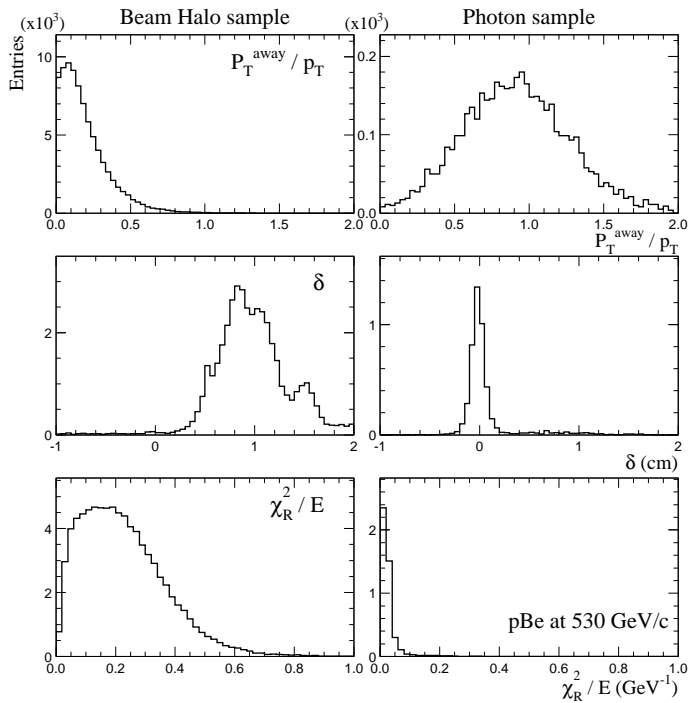}
\caption{Distributions in $P_T^{away}/ p_T $, $\delta$, and  
$\chi^2_R/E$ for showers from beam-halo muons (left) and photons 
(right) with $p_T >
5$~GeV/$c$. The beam-halo muon sample was obtained by requiring
signals from the VW quadrant shadowing the reconstructed shower. The
photon sample was obtained by imposing all the standard  
beam-halo muon rejection
criteria with the exception of the requirement on the variable plotted.}
\label{fig:muons}
\end{figure}

Distributions of $P_T^{away}/p_T$, $\delta$, and $\chi^2_R/E$ are
shown in Fig.~\ref{fig:muons} for a beam-halo muon enriched sample
(left) and a photon enriched sample (right).  The rejection
criteria discussed above suppressed beam-halo muon-induced background 
while having little impact upon the signal.  The effect of
these rejection criteria on the physics signal was checked using more
restrictive selection criteria.  The fraction of signal lost by the
application of each of the muon-rejection criteria provided a
measure of the signal efficiency of these criteria.
The application of the muon rejection criteria is estimated
to result in the loss of $\approx$7\% of the direct photons 
at $p_T= 4$~GeV/$c$, and $\approx$4\% at $p_T =
7$~GeV/$c$, for the 530~and 800~GeV/$c$ beam data.  The 
corresponding losses for the 515~GeV/$c$ $\pi^-$ beam data
are estimated to be 12\% and 8\%
at $p_T=4$ and $7$~GeV/$c$,
respectively.  Corrections for these losses, which are functions
of $p_T$ and \ycm, have been included in the results
reported in this paper.

\subsection{Direct-photon candidates}

\begin{figure}[t]
\epsfxsize=\figsize
\epsfbox[\bbcoord]{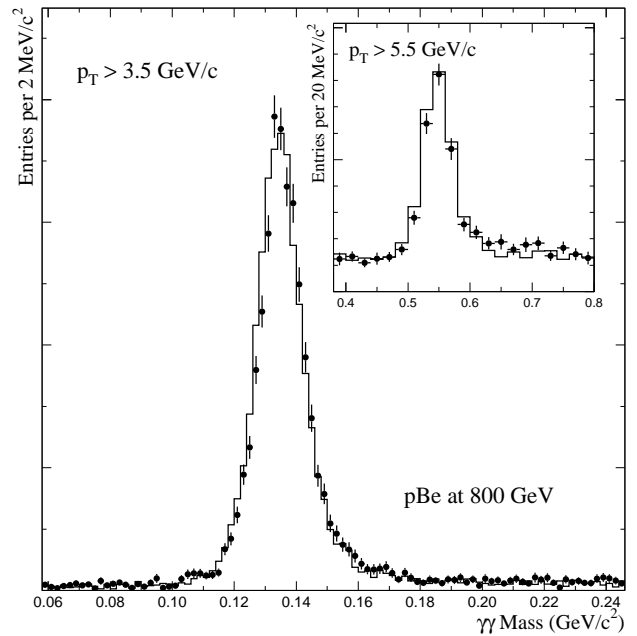}
\caption{$\gamma\gamma$ mass distributions in the $\pi^0$ and $\eta$
(inset) mass regions for photon pairs with
$A_{\gamma\gamma}\leq0.75$ and that satisfy the
specified minimum $p_T$ requirement in the  
data (histograms) compared to the results of the
detailed detector simulation ($\bullet$). 
The distributions have been normalized to the same area.}
\label{fig:mc-pi0-eta}
\end{figure}

\begin{figure}[t]
\epsfxsize=\figsize
\epsfbox[\bbcoord]{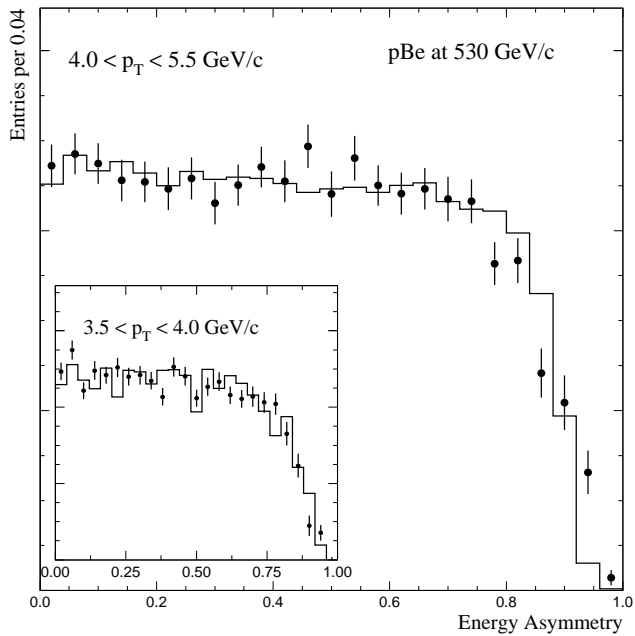}
\caption{Comparison of the side-band subtracted
energy asymmetry distribution 
for photons from reconstructed $\pi^0$ decays in data
(histogram) and the detailed detector simulation ($\bullet$).
Comparisons for two $p_T$ intervals, $4.0<p_T<5.5$~GeV/$c$ and
$3.5<p_T<4.0$~GeV/$c$ (inset) are shown.  The distributions have been
normalized to the same area.}
\label{fig:mc-A}
\end{figure}

The most significant source of direct-photon background is the
electromagnetic decay of hadrons such as $\pi^0$'s and $\eta$'s.
Photons were rejected from the direct-photon
candidate sample if they combined with
another photon in the same EMLAC octant to form a $\gamma\gamma$ pair
with invariant mass, $M_{\gamma\gamma}$, in the $\pi^0$ or $\eta$ mass
region and with energy asymmetry, $A_{\gamma\gamma} = |E
_{\gamma_1}-E_{\gamma_2}|/ (E_{\gamma_1}+ E_{\gamma_2})$, less than
the value specified for each direct-photon candidate definition
(described below).  The residual background from these mesons, as well
as from other sources of background, was evaluated using Monte Carlo
samples that contained no generated direct photons.

In the neutral meson
analyses~\cite{E706-pos-pieta,E706-neg-pieta},
a $\pi^0$ candidate was defined as a combination of two
photons with $A_{\gamma\gamma}\leq0.75$
detected within the fiducial region of the same EMLAC octant
with $100 < M_{\gamma\gamma} < 180~{\rm MeV}/c^2$;
$\eta$ candidates were similarly defined with $450 < M_{\gamma\gamma} <
650~{\rm MeV}/c^2$ (Fig.~\ref{fig:mc-pi0-eta}).  
The energy asymmetry requirement reduces
uncertainties due to 
low energy photons.
Background subtracted energy asymmetry distributions for photons in the
$\pi^0$ mass region are shown in Fig.~\ref{fig:mc-A}. 
Combinatorial backgrounds under the peak regions 
in the mass distributions were
evaluated using sideband subtraction.  

Three direct-photon candidate definitions were used in this analysis:
$75N$, $90N$, and $75S$, which differed in the treatment of the
background due to the decays of $\pi^0$ mesons (the largest source of
background).  Corrections and backgrounds were calculated
independently for each definition.  In the $75N$ ($90N$) definition,
any photon that formed a combination with another photon in the same
EMLAC octant with $M_{\gamma\gamma}$ in the $\pi^0$~mass region and
$A_{\gamma\gamma}\leq0.75$ ($A_{\gamma\gamma}\leq0.9$), was eliminated from
the direct-photon candidate sample. 
A photon was also rejected if it formed a
$\gamma\gamma$~pair with another photon in the same octant with
invariant mass in the $\eta$~mass region with $A_{\gamma\gamma}\leq0.8$.

The third definition, $75S$, rejected the same photons as the $75N$
definition.  However, photons that formed mass combinations with
other photons in the $\pi^0$~and $\eta$~sideband regions were included
with a weight of~2 (when these photons were in the same octant)
to explicitly account for direct-photon losses due to
accidental combinations under the $\pi^0$ and $\eta$ mass peaks.
The $75N$ and $90N$ results both rely on the detailed simulation to account
for the accidental losses of direct photons due to mass combinations
with other photons that happen to land in the $\pi^0$ or $\eta$~mass
windows.

Each of these definitions have relative strengths and weaknesses. The
$90N$ definition rejects the most background, however, the residual
background in this definition is more sensitive to the simulation of the
low energy photons from 
$\pi^0$  decay, which cause the losses at high $A_{\gamma\gamma}$.
Additionally,
accidental direct-photon losses are largest for this definition and must be
accounted for by the simulation.  The $75N$ and $75S$ definitions are
less sensitive to the details of the low energy photon simulation, 
but lead to larger background levels.
Comparing the fully corrected results using three
definitions provided insights into the systematic uncertainties
associated with the direct-photon cross-section measurements.

\subsection{Detailed detector simulation}

The Meson West spectrometer was modeled using a detailed {\sc
geant}~\cite{geant} simulation (DGS).  The simulated particles
in the generated events
were propagated through the {\sc geant} model of the 
spectrometer and interacted appropriately.  Simulated showers in the
EMLAC were parameterized once the energies of the interacting
particles were
$<$10~MeV. A preprocessor converted {\sc
geant} information into the hits and strip energies simulating the
detailed response of the various detectors.  The preprocessor
simulated hardware effects such as channel noise and gain variations.

Single particle distributions, reconstructed data events, and the {\sc
herwig}~v5.6~\cite{herwig56} event generator were employed as
inputs to the detailed {\sc geant} simulation.  The direct-photon background
studies were carried out primarily using samples of {\sc herwig}
$2\rightarrow2$ QCD~hard parton scatters ({\tt IPROC=1500}).  No
direct photons were included in these samples.  
The detailed simulations were used to
evaluate contributions to the
direct-photon background from numerous sources
including 
$\pi^0$, $\eta$, $\omega$, $\eta^\prime$, $K^0_L$, $K^0_S$,
$e^{\pm}$, neutrons and charged pions.

Generated events were
processed through the reconstruction software used for the data
analysis.  This technique accounted for inefficiencies and biases in
the reconstruction algorithms.  
Additional information on the detailed
simulation of the Meson West spectrometer can be found 
elsewhere~\cite{E706-pos-pieta,apana}.

The simulated 
{\sc herwig} events were weighted (in an iterative fashion)
so that the resulting $\pi^0$ spectra agreed with our
measured cross
sections as functions of $p_T$ and \ycm~\cite{E706-pos-pieta,E706-neg-pieta}.
As a result, the final corrections were based on 
the shape of the spectra from data
rather than on the {\sc herwig} event generator~\cite{apana}.
Similarly, the generated $\eta$'s were weighted so that
the resulting simulated production spectra were consistent with our
measured $\eta/\pi^0$ results~\cite{note-etatopi}. 
{\sc herwig}~v5.6 was assumed to reproduce other
particle spectra after these were
weighted by the results of the fits 
to the $\pi^0$ data~\cite{note-herwig-omega,note-etaprime}.  
Photons reconstructed in
the background samples that satisfied the direct-photon candidate
requirements, \gb, were assigned the weighting appropriate to their
identified parent.

Figure~\ref{fig:mc-pi0-eta} shows $\gamma\gamma$~mass spectra in
the $\pi^0$ and $\eta$ mass regions 
compared to the results of the
detailed {\sc GEANT} simulation.  In addition to
demonstrating that the DGS simulated the EMLAC resolution well, the
agreement between the levels of combinatorial background indicates the
DGS provided a reasonable simulation of the underlying event
structure.  Figure~\ref{fig:mc-A} shows a comparison between the DGS
results and the data for the $\pi^0$ energy asymmetry after side-band
subtraction.  This figure illustrates that the simulation accurately
describes the losses of low energy photons.

\begin{figure}
\epsfxsize=\figsize
\epsfbox[\bbcoord]{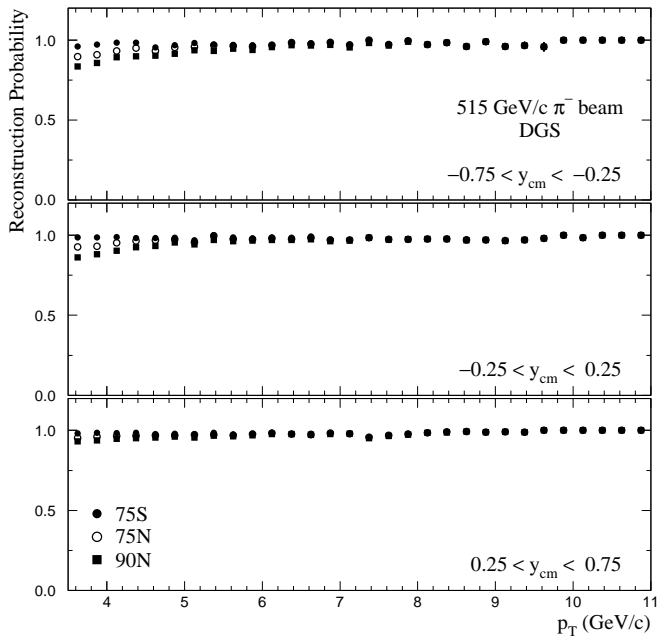}
\caption{Probabilities to reconstruct direct photons using the $75S$, $75N$, 
and $90N$ definitions as functions of generated $p_T$ of the
photons for several
rapidity ranges as determined via the detailed {\sc GEANT} simulation.}  
\label{fig:receff} 
\end{figure}

The direct-photon reconstruction
efficiency was evaluated using a dedicated sample of {\sc herwig}-generated
direct-photon events.  Reconstuction inefficiencies for direct photons
were relatively small over most of the kinematic range
(Fig.~\ref{fig:receff}).  

\subsection{EMLAC energy scale}

\begin{figure}
\epsfxsize=\figsize
\epsfbox[\bbcoordb]{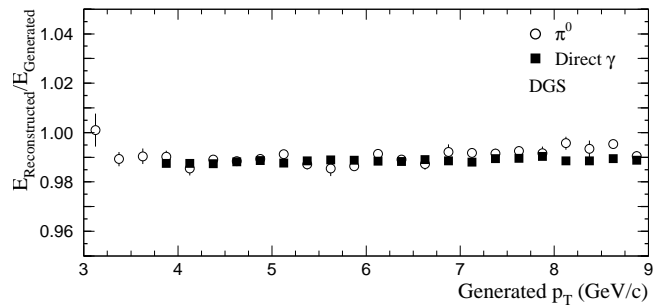}
\caption{Ratios of reconstructed to generated energy for $\pi^0$'s and 
direct photons as functions of their generated $p_T$ values
as determined via detailed {\sc GEANT} simulations.  The
small offset from 1.0 is due to the combined effects of the steeply
falling particle production spectra and the EMLAC resolution.}
\label{fig:dp-escale}
\end{figure}

The calibration of the EMLAC response was based on the
reconstructed masses of $\pi^0$ mesons in the $\gamma\gamma$ decay
mode and verified using $\eta$'s, $\omega$'s, converted photons
($e^+e^-$), and $J/\psi$'s~\cite{E706-calibration,begel}.  
The steeply falling $\pi^0$ $p_T$ spectrum, combined
with the calorimeter's resolution, results in a small offset
(1\%) in the mean reconstructed photon energies when plotted as
a function of the generated $p_T$
(Fig.~\ref{fig:dp-escale}).  To account for this offset and for
potential biases in the calibration procedure, the
simulated EMLAC was calibrated in the same manner as the real detector.  
The impact
of detector resolution on the energy scale calibration and on the
production spectra was incorporated into the Monte Carlo based
corrections~\cite{apana}.  The simulation was also employed to
evaluate the mean correction for energy deposited in the material
upstream of the EMLAC.  Two-photon mass distributions in the 
$\pi^0$ and $\eta$ mass regions from both data and DGS events
after energy scale calibration are illustrated in
Fig.~\ref{fig:mc-pi0-eta}.

The energy scale calibration was also verified to be appropriate for
direct photons.  This is important since the EMLAC energy response
was sensitive to event structure; direct photons are expected to be
more isolated than corresponding photons from meson decays since the
mesons are frequently
accompanied by other jet particles.  The single photon energy scale
was verified with $\eta\rightarrow\gamma\gamma$ where the photons were
widely separated, as well as with isolated electrons, and the DGS.
Figure~\ref{fig:dp-escale} displays a comparison of the ratio between
reconstructed and generated energy as functions of generated $p_T$
for $\pi^0$'s and direct photons from the DGS.  Cuts were applied to
both the photons from $\pi^0$ decay and direct photons 
ensuring equal treatment by the
reconstruction software~\cite{begel}.  These two distributions agree
well. 

\subsection{Direct-photon background}

\begin{figure}
\epsfxsize=\figsize
\epsfbox[\bbcoord]{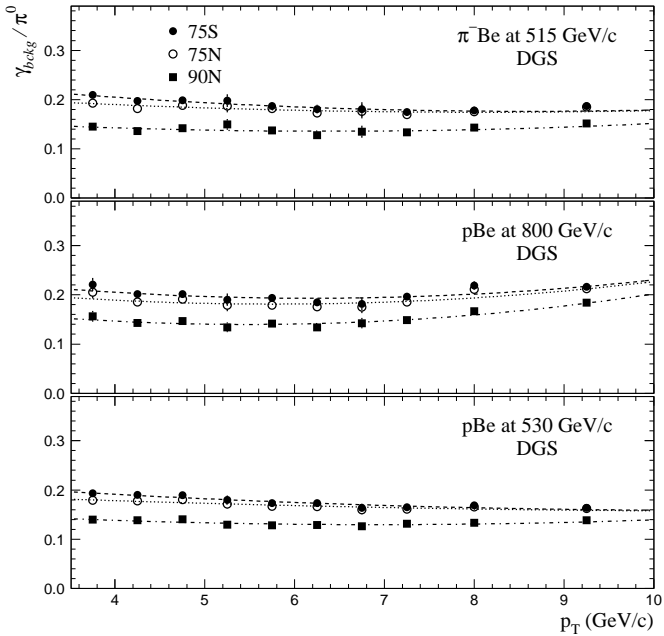}
\caption{\gb/$\pi^0$ for each direct-photon candidate definition 
as a function of $p_T$ based upon analysis of DGS samples excluding
the direct-photon samples. The points represent the binned results;
the curves represent 
fits to
these backgrounds integrated over rapidity.}
\label{fig:bckg_gpi} 
\end{figure}

The residual background to the direct-photon signal
from $\pi^0$'s and $\eta$'s, as well as from
other (previously mentioned) sources, was calculated by applying the 
$75N$, $90N$, and $75S$ photon
definitions to DGS samples that contained no generated direct photons.
Ratios of the resulting \gb\ to the measured $\pi^0$ cross
sections~\cite{E706-pos-pieta,E706-neg-pieta} are shown as functions
of $p_T$ for interactions 
on the beryllium target in Fig.~\ref{fig:bckg_gpi}.  
As indicated in the section describing the various direct-photon
candidate definitions, the $90N$ background
level is substantially lower than the $75N$ and $75S$ background levels.

\begin{figure}
\epsfxsize=\figsize
\epsfbox[\bbcoord]{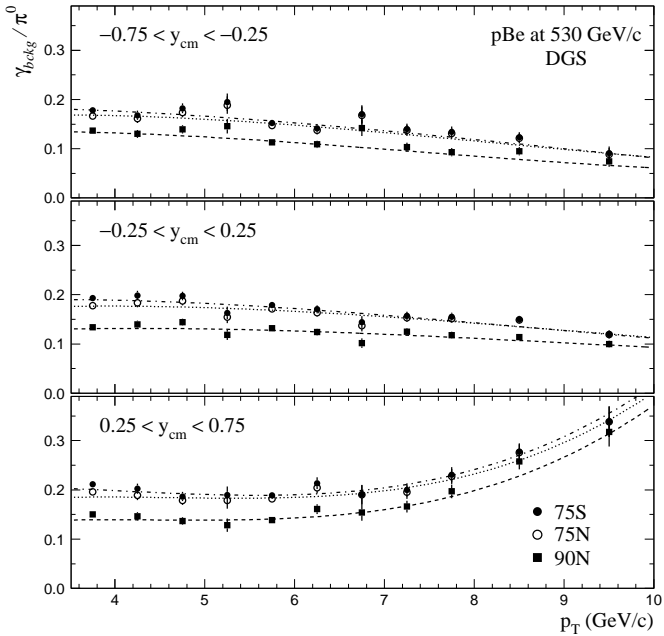}
\caption{\gb/$\pi^0$ for each direct-photon candidate definition
as a function of $p_T$ for backward, central, and forward rapidities
based upon analysis of DGS samples excluding the direct-photon samples.  The
curves represent the results of the background fits integrated over the
indicated rapidity range.}
\label{fig:bckg_gpi_rap}
\end{figure}

Figure~\ref{fig:bckg_gpi_rap} shows the
\gb/$\pi^0$ for several \ycm\ intervals for the simulated
530~GeV/$c$ $p$Be interactions. The shapes of the
background are quite similar in the backward and central \ycm\
regions.  However, at forward \ycm\ and high~$p_T$ the contribution
to the background due to
coalesced photons from $\pi^0$ decays is clearly visible.  The
\gb/$\pi^0$ were fitted to functions in $p_T$ and \ycm.  Projections
from the fits are indicated by the curves in
Figs.~\ref{fig:bckg_gpi} and~\ref{fig:bckg_gpi_rap}.

\begin{figure}
\epsfxsize=\figsize
\epsfbox[\bbcoord]{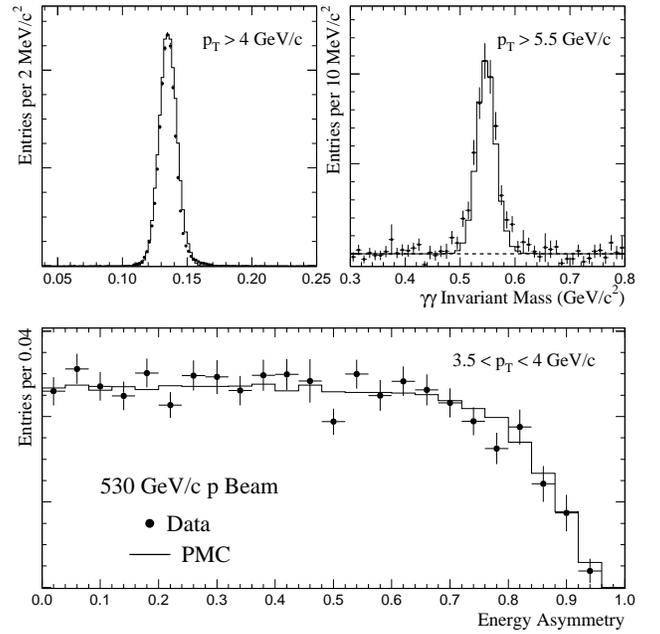}
\caption{Top: 
comparison between background-subtracted $\gamma\gamma$ mass distributions for
data~($\bullet$) and the PMC (histogram) for photon pairs with
$A_{\gamma\gamma}\leq0.75$ 
in the $\pi^0$ and $\eta$ mass regions.
Note that the background levels are centered on zero (dashed line)
in the data mass
plots due to the background subtraction.  
Bottom:
comparison of the corresponding energy asymmetry distributions
for $\gamma$'s from $\pi^0$ decays.
The distributions have
been normalized to the same area.}
\label{fig:pmc}
\end{figure}

\subsection{Parameterized detector simulation}

A second Monte Carlo simulation of detector effects (PMC) was used to
cross check the DGS and for large statistics studies.  This simulation
employed parameterizations of physics cross sections and detector
responses~\cite{apana,begel}.  The inclusive $\pi^0$ and direct-photon
cross sections were parameterized as two-dimensional functions in $p_T$
and \ycm~\cite{apana}.  The $\eta$, $\omega$, and $\eta^\prime$ cross
sections were represented using values consistent with our measured
$\eta/\pi^0$~\cite{E706-pos-pieta,E706-neg-pieta,note-etatopi},
$\omega/\pi^0$~\cite{E706-omega}, and using 
$\eta^\prime/\pi^0$=0.85, as indicated in
Ref.~\cite{etaprimeratio} (after accounting for the  
updated $\eta^\prime\rightarrow\gamma\gamma$ branching
ratio~\cite{pdg}).

An interaction vertex was generated in the
simulated target for every event (Fig.~\ref{fig:vz}).  
This primary vertex
distribution was used to ensure the simulated photons
traversed the proper amount of material.  
Generated mesons were allowed to decay into final state particles.
Photons were
allowed to convert into $e^+e^-$ pairs when appropriate; the energy of
the resulting electrons was reduced using the {\sc geant} function for
bremsstrahlung radiation.  Electron four-vectors were smeared for
multiple scattering in the target region and the resolution of the tracking
system, and adjusted to account for the impulse due to the magnet.
The parameters of the photons that did not convert
were smeared to account for energy and position
resolution~\cite{E706-calibration}.  
Figure~{\ref{fig:pmc} displays a comparison between 
$\gamma\gamma$ mass distributions from the PMC and the
background-subtracted data in the $\pi^0$ and $\eta$ mass regions 
as well as the energy asymmetry distribution for photons
from $\pi^0$ decays.  
As illustrated by these plots, the PMC provides a reasonable
characterization of the data.

\begin{figure}
\epsfxsize=\figsize
\epsfbox[\bbcoord]{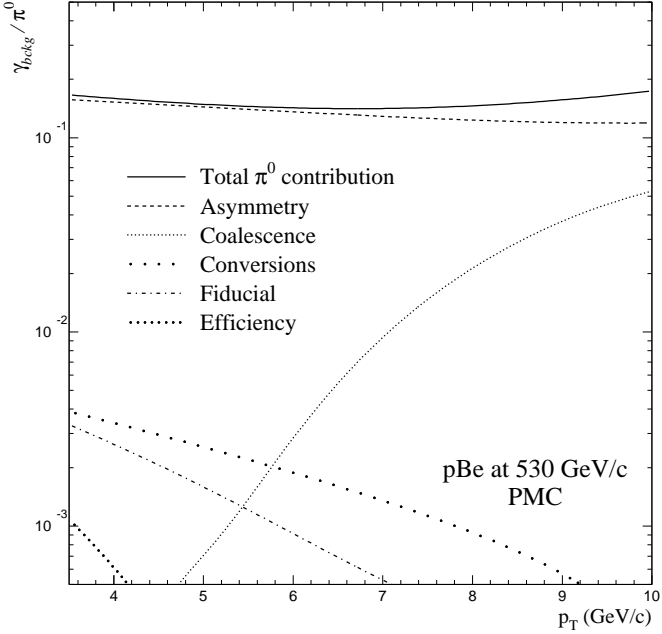}
\caption{
Contribution to \gb/$\pi^0$ from $\pi^0$ decays for the $75S$ photon
definition evaluated via the PMC.  Also shown are the contributions
from each of
the categories described in the text.}
\label{fig:pi0_bckg}
\end{figure}

The PMC was used to identify contributions to the direct-photon
background.  Photons from $\pi^0\longrightarrow\gamma\gamma$ decays
with $A_{\gamma\gamma}$ greater than required by the relevant photon
definition were the dominant
contribution to the background.  However, photons from $\pi^0$ decays
with smaller $A_{\gamma\gamma}$ also contribute to \gb\ when the
other photon is lost.  Hence, the remainder of the photons
from $\pi^0$ decays were classified according to our ability to detect
both photons from the $\pi^0$ decay.
The photons were projected to the front face of the EMLAC.  If one was
outside the EMLAC fiducial volume while the other was within
that fiducial volume, then the
latter contributed to \gb.  In addition, photons that landed in
different EMLAC octants were classified as background candidates.  Contributions
to \gb\ from photon conversions and detector inefficiencies were
obtained from the sample when both photons landed within a single
EMLAC octant.  Each photon was assigned a weight based on the
conversion probability of the other photon.  
The \gb\ contribution from
detector inefficiencies was determined by weighting each photon by the
non-detection probability for the other photon.  Double counting
between these sources was eliminated by applying the non-conversion
probability of the undetected photon to each \gb\ contribution.  
Since the \gb\ were
required to be inside the fiducial volume, the impact of
the EMLAC acceptance was properly taken into account.

The DGS was used to assess the coalescence probability function as the
coalescence of two photons from a $\pi^0$ decay into a single
reconstructed shower depended on many factors including the relative
geometry of the two photons, their total and relative energies, and
the ability of the reconstructor to resolve the photons.  Background
contributions from this source were assigned the $\pi^0$ energy and
weighted by the non-conversion probability and detection efficiency of
each photon and the acceptance of the higher energy photon.

The ratio of \gb\ from $\pi^0\rightarrow\gamma\gamma$ decays to
generated $\pi^0$'s is shown in Fig.~\ref{fig:pi0_bckg}
for incident 530~GeV/$c$ $p$ beam on the beryllium targets.  
The contributions from each of the categories described above
are also shown in the same figure. It is
evident from the figure that photons resulting from highly asymmetric
$\pi^0$ decays are the dominant contribution to the direct-photon
background, although, at very high $p_T$, the coalescence contribution
is also significant.

The background contribution from the decay
$\eta\rightarrow\gamma\gamma$ was evaluated in an analogous manner.
Photons from $\eta$ decays with
energy asymmetry $A_{\gamma\gamma}>0.8$ were included in 
the direct-photon background.
Photons from the decays
$\omega\rightarrow\pi^0\gamma$, $\eta^\prime\rightarrow\rho\gamma$,
$\eta^\prime\rightarrow\omega\gamma$,
$\eta^\prime\rightarrow\gamma\gamma$,
and $\pi^0\rightarrow\gamma e^+e^-$ were also evaluated
as potential contributors to \gb.
Photons arising from the subsequent decay of the $\pi^0$
in the case of the $\omega$ decay and the $\rho\rightarrow\pi^0\pi^0$ decays
were not considered since those contributions were already included in
the simulation of the inclusive $\pi^0$ sample.  
The background contribution from the
$e^+e^-$ pair in $\pi^0\rightarrow\gamma e^+e^-$ decays was negligible
due to the small branching ratio for this decay mode combined with 
the typically
significantly lower $p_T$ of the daughter electrons relative to
the parent $\pi^0$. 
Furthermore, most electrons were rejected by the distance-to-nearest-track
requirement (Fig.~\ref{fig:dtrk}).

\begin{figure}
\epsfxsize=\figsize
\epsfbox[\bbcoord]{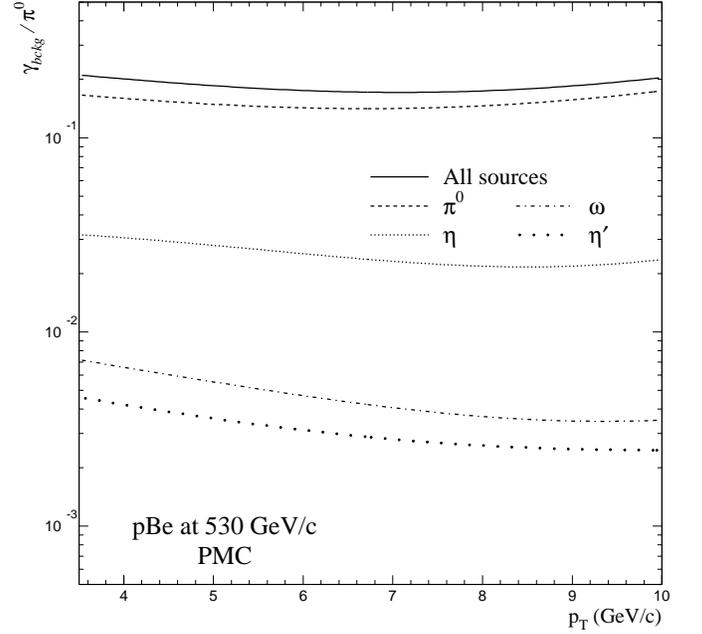}
\caption{
Contribution to \gb/$\pi^0$ from $\pi^0$'s, $\eta$'s, $\omega$'s, and
$\eta^\prime$'s for the $75S$ photon definition as evaluated
using the PMC.}
\label{fig:pi0_bckg_source}
\end{figure}

The total \gb/$\pi^0$ as determined using the PMC is shown in
Fig.~\ref{fig:pi0_bckg_source} for incident 530~GeV/$c$ $p$ on
beryllium. 
The contribution to 
\gb/$\pi^0$ due to each of the particles considered in the
PMC is also shown in that figure. 
The contribution from $\eta$ decays is roughly 20\% of the
$\pi^0$ contribution, as expected from simple considerations of the $\eta$ and
$\pi^0$ production rates~\cite{E706-pos-pieta,E706-neg-pieta} and
their respective two-photon branching ratios~\cite{pdg}.

\begin{figure}
\epsfxsize=\figsize
\epsfbox[\bbcoordb]{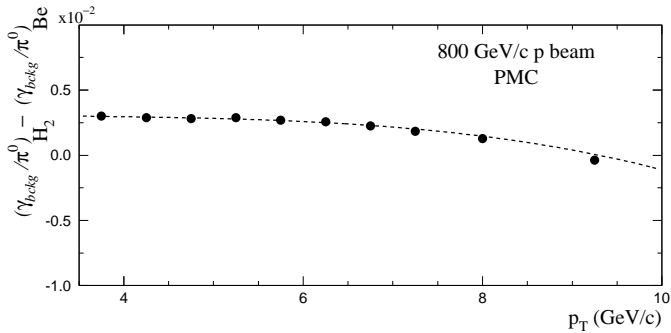}
\caption{Difference between 
\gb/$\pi^0$ for the hydrogen and beryllium targets 
as a function of $p_T$ as determined via the PMC. 
The dashed curve represents a fit to this difference.}
\label{fig:tgt_fix}
\end{figure}

\subsection{Direct-photon background for hydrogen target data}

Fits to \gb/$\pi^0$ evaluated using the DGS
were only made for the beryllium target due to
relatively poor DGS statistics in the other targets.  However,
\gb/$\pi^0$ is expected to be slightly different for each target
due to the different amounts of target material the photons must
traverse. A correction to \gb/$\pi^0$ was therefore evaluated using
the PMC. The difference between \gb/$\pi^0$ on the hydrogen and beryllium
targets is shown for the 800~GeV/$c$ proton beam sample in
Fig.~\ref{fig:tgt_fix}.  The corresponding results are similar
for the other incident beams.  Note
that the difference is small ($<0.003$) but positive at low $p_T$ 
because the conversion
background contribution is larger for the more upstream hydrogen
target.  However, at high $p_T$, the correction becomes 
very slightly negative
because the contribution due to coalescence is reduced by the 
additional photon
conversions.  The target differences were fit as functions of $p_T$
and \ycm\ for each incident beam and applied as additive corrections
to the nominal \gb/$\pi^0$ fit obtained using the DGS and the
beryllium target.

\begin{figure}
\epsfxsize=\figsize
\epsfbox[\bbcoord]{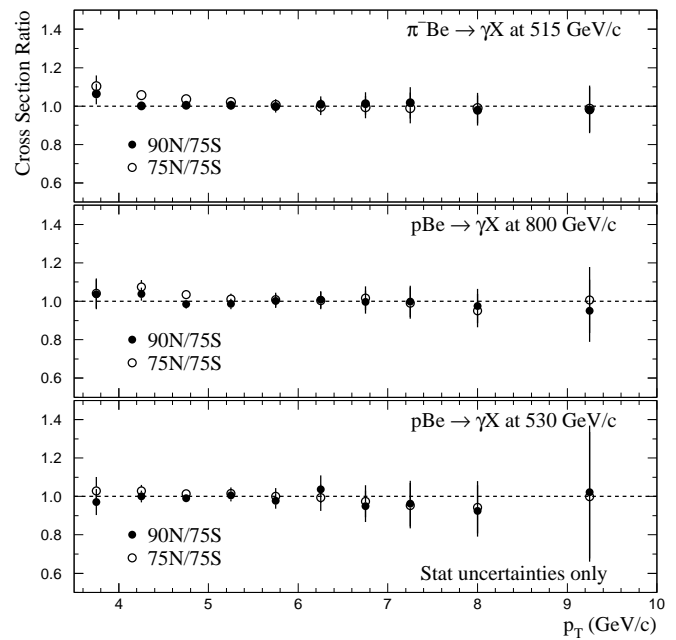}
\caption{
Ratios between direct-photon cross sections evaluated
using the $90N$ and $75S$ direct-photon candidate definitions 
as functions of photon $p_T$ for the three major incident beams.
Ratios between direct-photon cross sections evaluated
using the $75N$ and $75S$ direct-photon candidate definitions
are also shown.}
\label{fig:sys_def}
\end{figure}

\subsection{Background subtraction uncertainties}

The systematic uncertainty in the background subtraction was
estimated via the sensitivity of the results to the various
direct-photon definitions and to 
variations in the size of the
$\pi^0$ and $\eta$ side-band regions and via detailed studies of the
Monte Carlo simulations.  Ratios of direct-photon 
cross sections obtained using the
different candidate definitions are displayed in
Fig.~\ref{fig:sys_def} for all three incident-beam samples.
These ratios generally differed from unity by
$<5$\%. This very good agreement between the direct-photon
cross sections measured using these different
definitions illustrates the 
robust nature of our results.

\begin{figure}
\epsfxsize=\figsize
\epsfbox[\bbcoord]{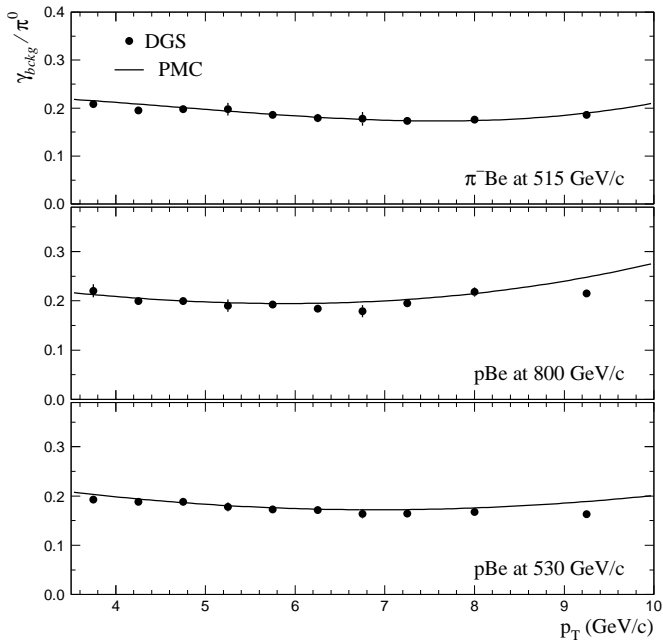}
\caption{
Comparison of \gb/$\pi^0$ between the 
restricted background source DGS~($\bullet$) 
and the PMC
(curve) for each of the three major incident beam configurations
for the $75S$ photon definition.}
\label{fig:gpi_hw_pmc}
\end{figure}

\begin{figure}
\epsfxsize=\figsize
\epsfbox[\bbcoord]{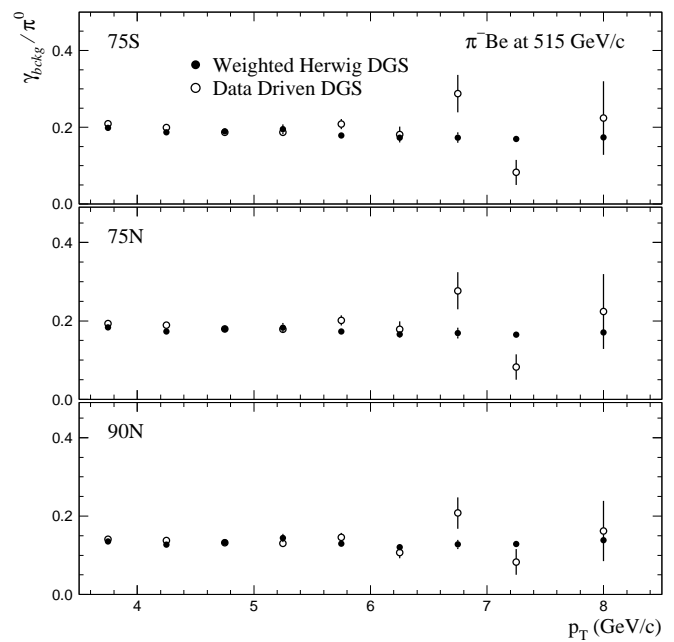}
\caption{
Comparison of \gb/$\pi^0$ from the DGS using weighted {\sc herwig} 
events and using
reconstructed data inputs as the event generators
for the $75S$, $75N$, and $90N$ direct-photon candidate definitions.}
\label{fig:gpi_dd_hw}
\end{figure}

Comparisons between detailed and parameterized detector simulations
provided increased confidence in our understanding of the direct-photon
background since the two simulations represent essentially
independent calculations of \gb.  The PMC-evaluated \gb\ is based on
relatively simple considerations and included only the major
background sources.  Special runs of the DGS were performed limiting
the background sources to those included in the PMC. \gb/$\pi^0$ from
the restricted background source DGS and the PMC are compared in
Fig.~\ref{fig:gpi_hw_pmc}.  The two simulation results agree quite well.

Measured four-vectors from particles reconstructed in data
events were also employed as inputs to the {\sc geant} simulation.
This provided an additional measure of the uncertainty associated with
the modeling of the detector environment.  A comparison of \gb/$\pi^0$
produced in the DGS using the weighted {\sc herwig} event generator
and reconstructed data events is shown in Fig.~\ref{fig:gpi_dd_hw}.
Only background sources common to both simulations were included in
the comparison.  The \gb/$\pi^0$ results evaluated via these two
methods are consistent.

The impact of uncertainty in $\eta/\pi^0$ on \gb/$\pi^0$ was
determined using the PMC.  The values used for $\eta/\pi^0$ were shifted by
$\pm 0.04$;
these
shifts resulted in a $\pm 1.5\%$ change in \gb/$\pi^0$.

\subsection{Normalization}

Electronic scalers that counted signals from the beam hodoscope and
beam hole counters were used to establish the number of beam particles
incident on the target.  Other scalers logged the state of the trigger
and data acquisition system.  Information from these scalers was used
to determine the number of beam particles incident upon the
spectrometer when it was ready to record data. Small corrections for
multiple occupancy in the beam hodoscope and for absorption of beam in
the target material were included in the normalization.

Additional discussion on the normalization including details regarding
independent normalization methods and cross checks can be found
elsewhere~\cite{E706-pos-pieta,E706-neg-pieta}.  
The net systematic uncertainty in the
overall normalization is 8\%.
This evaluation is based upon several considerations:
the good
agreement between results from the independent normalization
methods,  the stability of the cross section results from
various sections of the data,
the internal consistency
of the scalers, and a detailed analysis of the design, implementation,
and performance of the trigger. 

\subsection{Secondary beam contamination}

The Meson West beamline transported 0.8~TeV/$c$ primary protons from
the Fermilab Tevatron and 0.5~TeV/$c$ secondary beams of both
polarities. 
The 800~GeV/$c$ primary
beam has a negligible momentum spread.
The negative secondary pion beam was determined to have a mean momentum of
$515~\pm~2~{\rm GeV}/c$ with an estimated halfwidth of 
30~GeV/$c$.  The secondary proton beam was determined to have a
mean momentum of $530~\pm~2~{\rm GeV}/c$ and similar momentum spread.
This momentum spread in the secondary beams introduces 
a small uncertainty ($<4\%$) in
comparisons with theory evaluated at the mean beam momentum.

The minority particle fractions in the 530~GeV/$c$
positive secondary beam were measured using the beamline \v{C}erenkov detector
to be 2.75\% $\pi^+$ and 0.5\% $K^+$~\cite{striley}.  Analysis of the
515~GeV/$c$ negative secondary beam indicated 1\% $K^-$ component,
and negligible $\bar{p}$ content.

The direct-photon cross sections for incident 530~GeV/$c$ $p$ beam
have been corrected for the small incident pion contamination as follows.
An estimate of the direct-photon cross section from incident $\pi^+$
was generated from a product of our measured incident $\pi^-$ 
cross sections and 
the ratio of next-to-leading-order PQCD 
calculations~\cite{NLO-aurenche,NLO-berger}
for direct-photon production in $\pi^+$ and $\pi^-$ beams
to account for the differences in incident quark content.
The direct-photon cross sections reported here for
incident 530~GeV/$c$ protons were  
corrected to account for the minority pions using this estimate
of the direct-photon cross section from incident $\pi^+$ combined with the
measured incident minority particle fraction.
The correction was 
$<1\%$ at $p_T=4$~GeV/$c$ and $<5$\% at
$p_T=8$~GeV/$c$.  

The direct-photon cross-section measurements have not been corrected
for the incident kaon contributions due to
the lack of an incident kaon PDF.  The potential bias from
this source was studied using restrictive cuts based upon the
\v{C}erenkov detector data and the impact of the kaon 
contamination upon the reported cross-section measurements
is expected to be negligible.

\begin{figure}
\epsfxsize=\figsize
\epsfbox[\bbcoord]{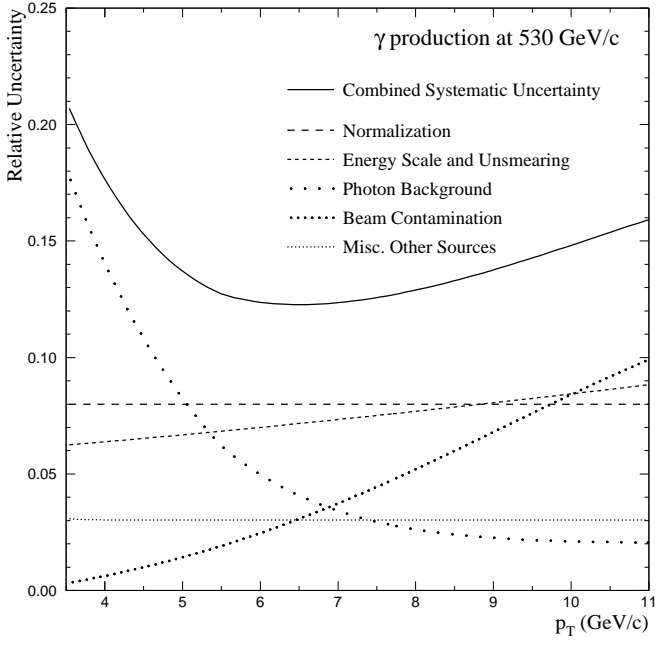}
\caption{Relative systematic 
uncertainty for the direct-photon cross section as a function of $p_T$
for incident 530~GeV/$c$ protons.}
\label{fig:sysunc}
\end{figure}

\subsection{Summary of systematic uncertainties}

The principal contributions to the systematic uncertainty 
in the direct-photon cross-section results arose from
the following sources: normalization, background subtraction, 
the calibration of photon energy response and
detector-resolution unsmearing, and incident
beam contamination (for the 530~GeV/$c$ $p$ beam).
Other sources of uncertainty, which 
contributed at the one to two percent level, included: photon
reconstruction, beam-halo muon rejection, geometric acceptance, photon
conversions, trigger response, and vertex finding.  The relative systematic
uncertainty in the direct-photon cross section
is presented as a function of $p_T$ in
Fig.~\ref{fig:sysunc} for the 530~GeV/$c$ $p$Be data.
With the exception of the beam contamination contribution, the
corresponding uncertainties for the other measurements are similar.
The total systematic uncertainties, combined in quadrature, are quoted
with the cross sections in the appropriate tables.  Note that some of these
contributions to the systematic uncertainty (e.g. normalization) are
strongly correlated between bins.


\begin{figure}
\epsfxsize=\figsize
\epsfbox[\bbcoord]{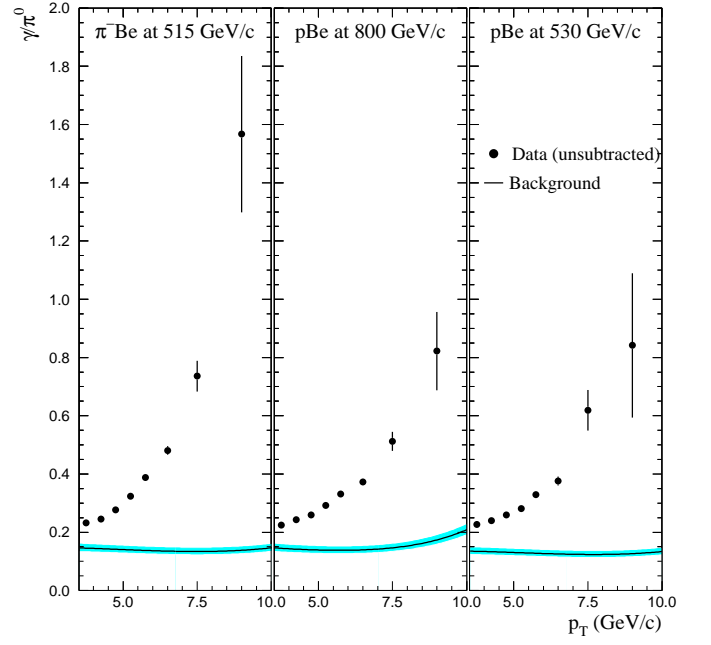}
\caption{
Ratios of the $90N$ direct-photon candidate spectra to the measured
$\pi^0$ cross sections (points) compared to \gb/$\pi^0$ from the DGS
(curves) as functions of $p_T$ for the data samples considered in this
paper.  The error bars represent statistical contributions to the 
uncertainties.
The width of each band around the background represents the systematic
uncertainty on that background.}
\label{fig:gpi}
\end{figure}

\begin{figure}
\epsfxsize=\figsize
\epsfbox[\bbcoord]{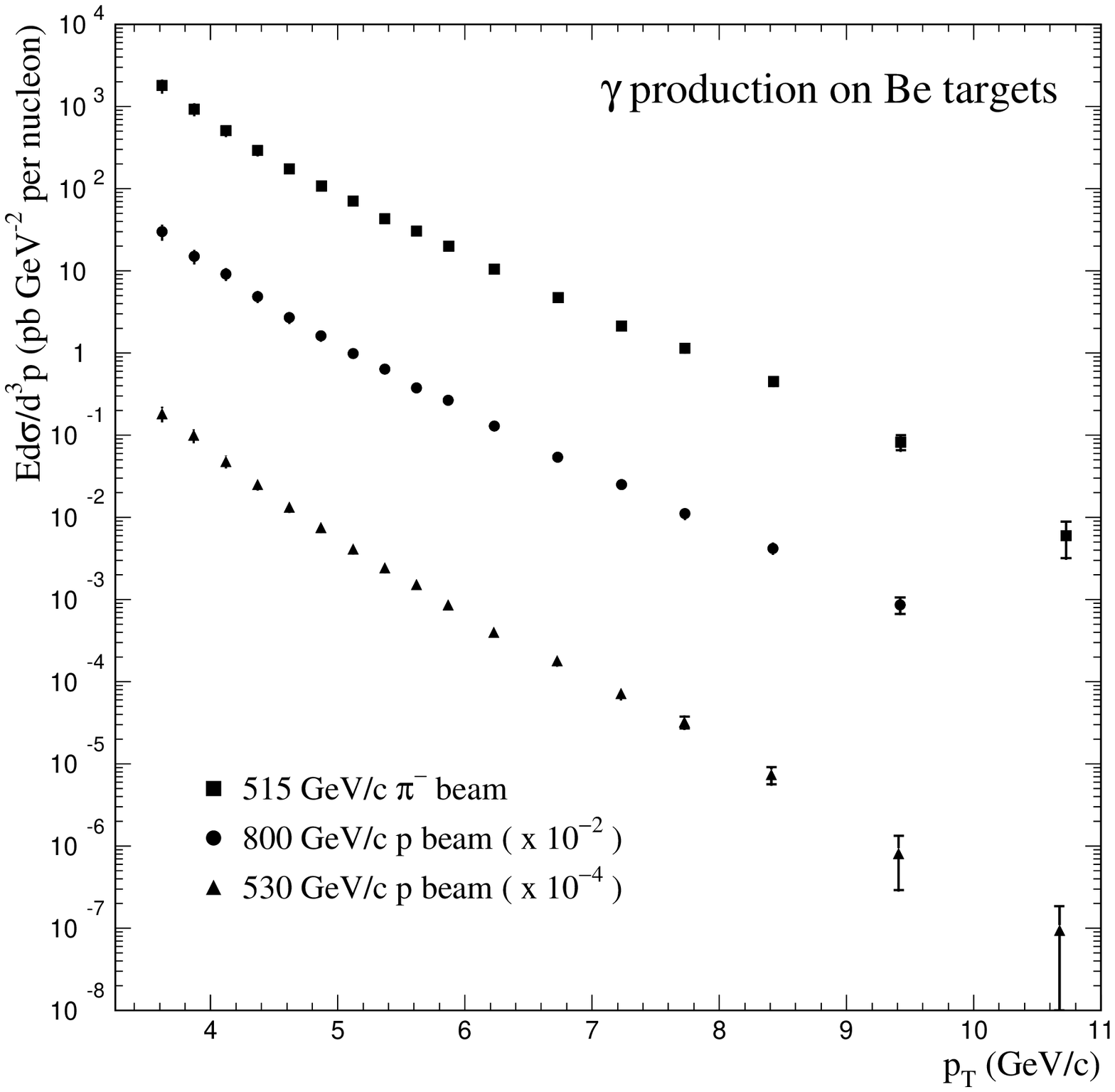}
\caption{
Invariant differential cross sections per nucleon for direct-photon
production as functions of $p_T$, averaged over rapidity, for
515~GeV/$c$ $\pi^-$ and 800 and 530~GeV/$c$ proton beams incident upon
beryllium. 
The error bars represent the statistical and systematic uncertainties
combined in quadrature; the innermost interval 
indicates the statistical uncertainties.}
\label{fig:xs_pt_all}
\end{figure}

\begin{figure}
\epsfxsize=\figsize
\epsfbox[\bbcoord]{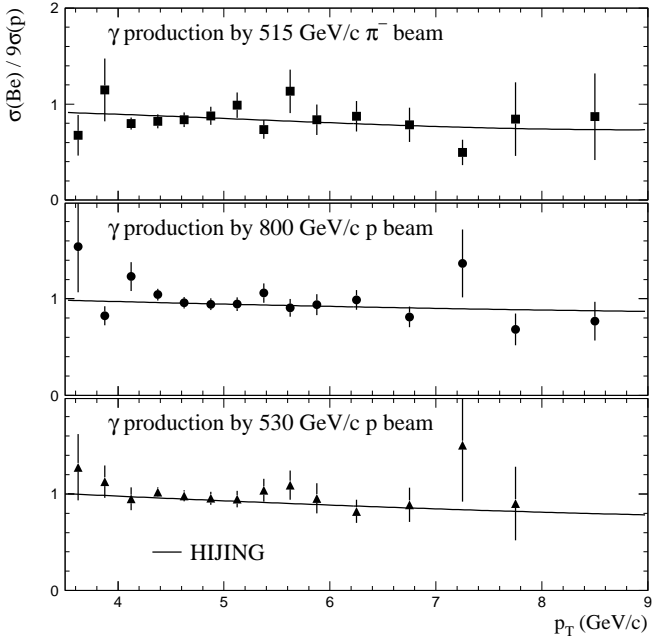}
\caption{Ratios of inclusive direct-photon production 
per nucleon in $p$Be to those in $pp$ collisions as functions
of photon $p_T$, compared with curves representing 
predictions from {\sc hijing}. The error bars
represent only statistical contributions to the
measurement uncertainties.
\label{fig:comp_hijing}}
\end{figure}

\begin{figure}
\epsfxsize=\figsize
\epsfbox[\bbcoord]{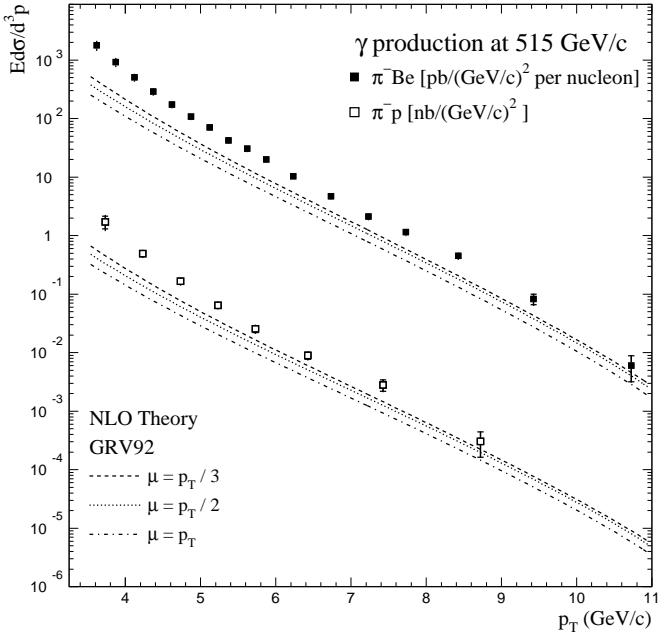}
\caption{
Invariant differential cross sections per nucleon for direct-photon
production as functions of $p_T$ in $\pi^-$Be and $\pi^-p$
interactions at 515~GeV/$c$. 
The error bars represent the statistical and systematic uncertainties
combined in quadrature; the innermost interval 
indicates the statistical uncertainties.
Overlaid on the data are NLO PQCD
results for three choices of the factorization and renormalization scales.}
\label{fig:xs_pt_515}
\end{figure}

\begin{figure}
\epsfxsize=\figsize
\epsfbox[\bbcoord]{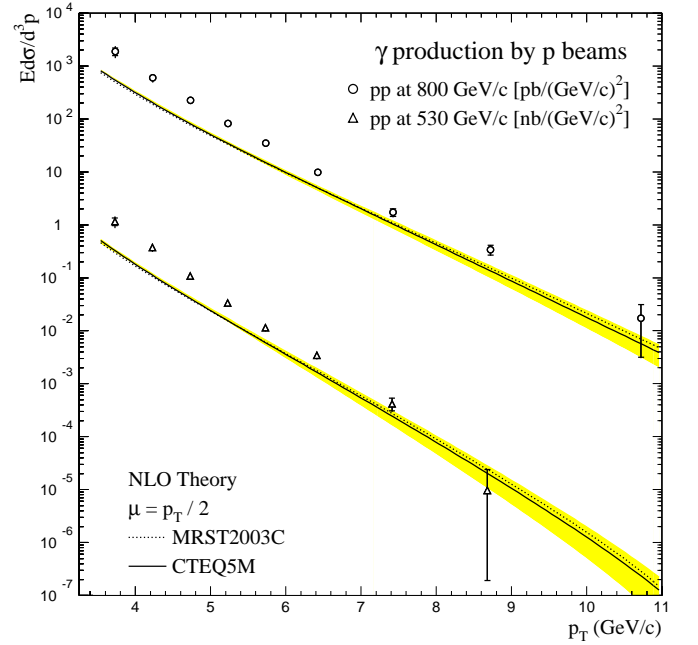}
\caption{
Invariant differential cross sections per nucleon for direct-photon
production as functions of $p_T$ in $pp$ interactions at 800 and
530~GeV/$c$.  
The error bars represent the statistical and systematic uncertainties
combined in quadrature; the innermost interval 
indicates the statistical uncertainties.
Overlaid on the data are NLO PQCD predictions for the CTEQ5M and
MRST2003C PDF; the band illustrates the PDF uncertainty estimated via
the MRST2001E set.}
\label{fig:xs_pt_530_800}
\end{figure}

\section{Results and Discussion}

\subsection{Results}

The ratios of the $90N$ direct-photon candidate
spectra to the measured $\pi^0$ cross
sections~\cite{E706-pos-pieta,E706-neg-pieta} as functions of $p_T$
are compared to \gb/$\pi^0$ from the DGS in Fig.~\ref{fig:gpi}.
All three samples are characterized by rapidly rising direct-photon
signal to background ratios.

The direct-photon cross sections were extracted statistically using
the fits to \gb/$\pi^0$.   
The invariant differential cross sections per nucleon for
direct-photon production from 515~GeV/$c$ $\pi^-$ and 800 and
530~GeV/$c$ proton beams incident on beryllium are presented as
functions of $p_T$ in Fig.~\ref{fig:xs_pt_all}.  Results from
515~GeV/$c$ $\pi^-$ and 530~GeV/$c$ $p$ beams are averaged over the
rapidity range $-0.75\le\ycm\le 0.75$; results from the 800~GeV/$c$
$p$ beam are averaged over $-1.0\le\ycm\le 0.5$.  Data points are
plotted at abscissas that correspond to the average value of the
cross section in each $p_T$ bin, assuming local exponential dependence
on $p_T$~\cite{laff}.  The inclusive cross sections are tabulated in
the Appendix (Tables~\ref{table_xs_pt_be}---\ref{table_xs_pt_rap_515_p}).  

Ratios of the direct-photon cross sections 
per nucleon on beryllium to those on
hydrogen are shown versus $p_T$ in Fig.~\ref{fig:comp_hijing}.  The
ratios are compared with results from the Monte Carlo program {\sc
hijing}~\cite{wang-note}.
The results from {\sc hijing}, which is designed to simulate particle
production in $pp$, $pA$, and $AA$ collisions,  
are in good agreement with the data.  The PQCD
calculations described below for interactions on 
beryllium have been adjusted to
account for nuclear effects using {\sc hijing} results.

\begin{figure}
\epsfxsize=\figsize
\epsfbox[\bbcoord]{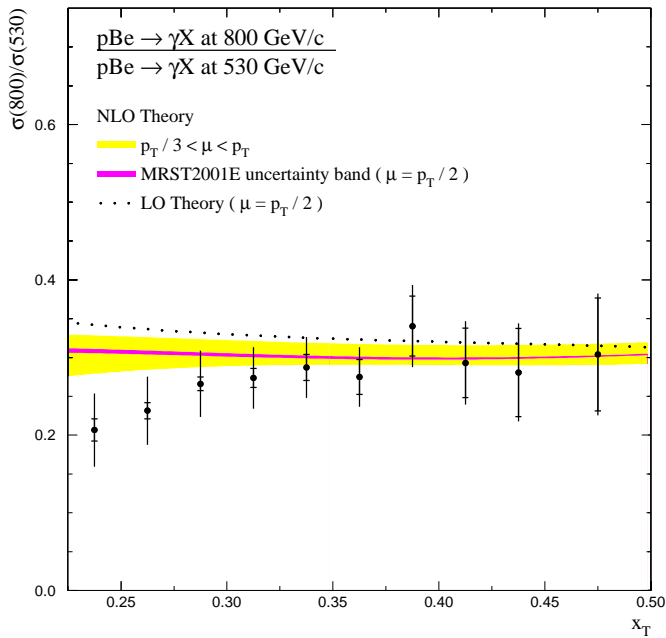}
\caption{The ratio of 
800~GeV/$c$ to 530~GeV/$c$ proton beam direct-photon cross sections
as a function of $x_T$.  
The error bars represent the statistical and systematic uncertainties
combined in quadrature; the innermost interval 
indicates the statistical uncertainties.
The dotted curve shows the results of
LO PQCD calculations using the CTEQ5L PDF and $\mu=p_T/2$.
Shaded bands illustrate the results from NLO PQCD calculations. 
The outer band is bounded by $\mu =
p_T/3$ and $\mu = p_T$; the inner band by the uncertainty of the MRST2001E
PDF set.
\label{fig:eng_dep}}
\end{figure}

\begin{figure}
\epsfxsize=\figsize
\epsfbox[\bbcoord]{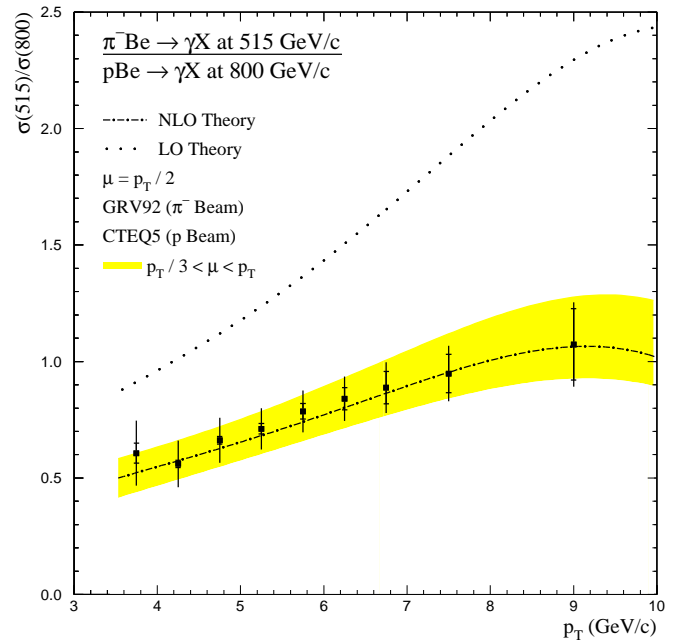}
\caption{The ratio of 
515~GeV/$c$ $\pi^-$ beam to 800~GeV/$c$ proton-beam direct-photon
cross sections as a function of $p_T$. 
The error bars represent the statistical and systematic uncertainties
combined in quadrature; the innermost interval 
indicates the statistical uncertainties.
Results of LO PQCD calculations using $\mu=p_T/2$ are represented
by the dotted curve.
The shaded band illustrates results from NLO PQCD calculations 
bounded by $\mu = p_T/3$ and
$\mu = p_T$.
\label{fig:beam_dep}}
\end{figure}

\begin{figure}
\epsfxsize=\figsize
\epsfbox[\bbcoord]{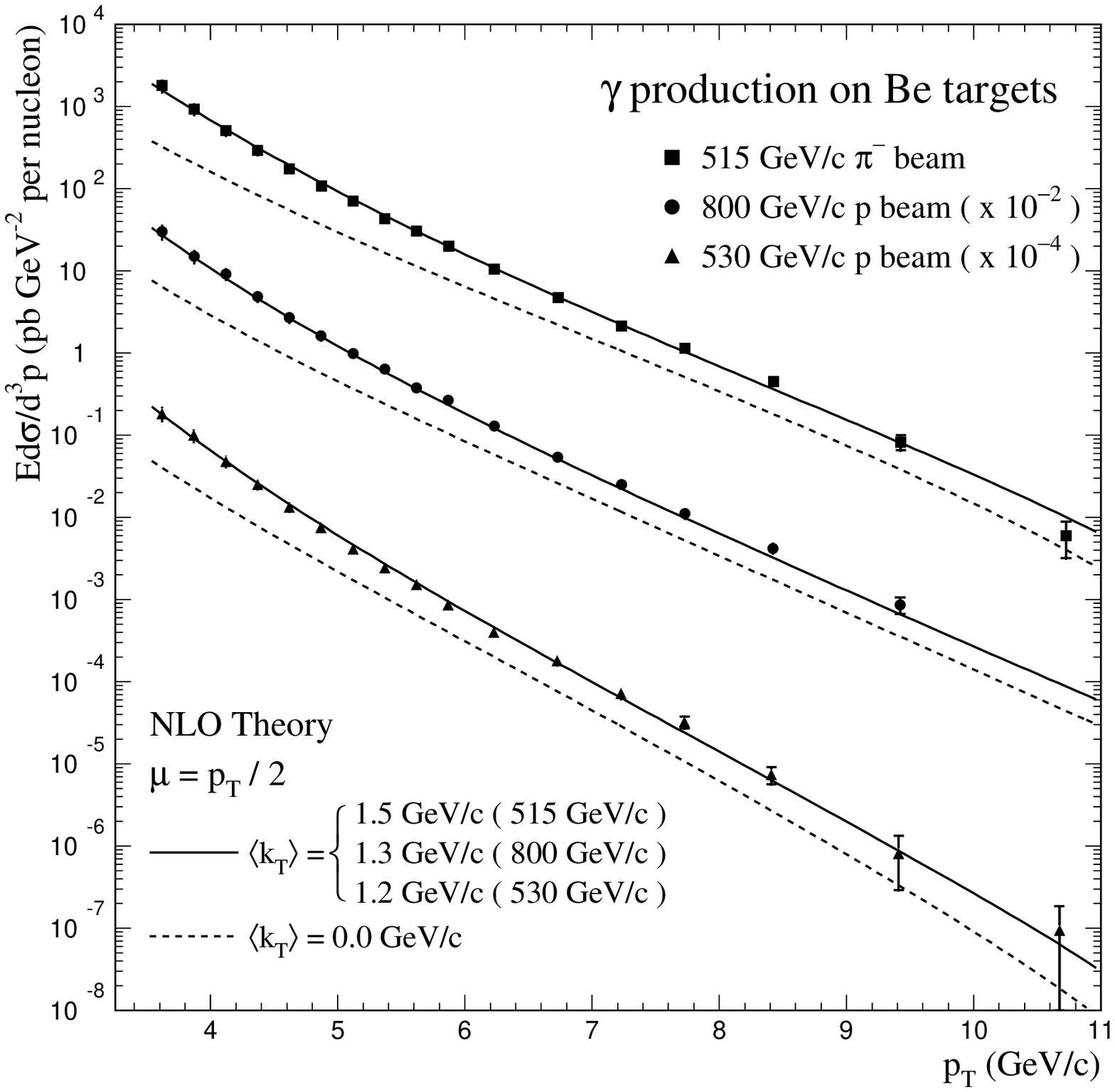}
\caption{
Invariant differential cross sections per nucleon for direct-photon
production as functions of $p_T$, averaged over rapidity, for
515~GeV/$c$ $\pi^-$ and 800 and 530~GeV/$c$ proton beams incident upon
beryllium.  
The error bars represent the statistical and systematic uncertainties
combined in quadrature; the innermost interval 
indicates the statistical uncertainties.
Overlaid on the data are NLO PQCD and
$k_T$-enhanced NLO PQCD calculations.  GRV92 PDF were used 
in the incident $\pi^-$ calculations, while
CTEQ5M PDF were used in the incident proton calculations.}
\label{fig:xs_pt_all_kt}
\end{figure}

\begin{figure}
\epsfxsize=\figsize
\epsfbox[\bbcoord]{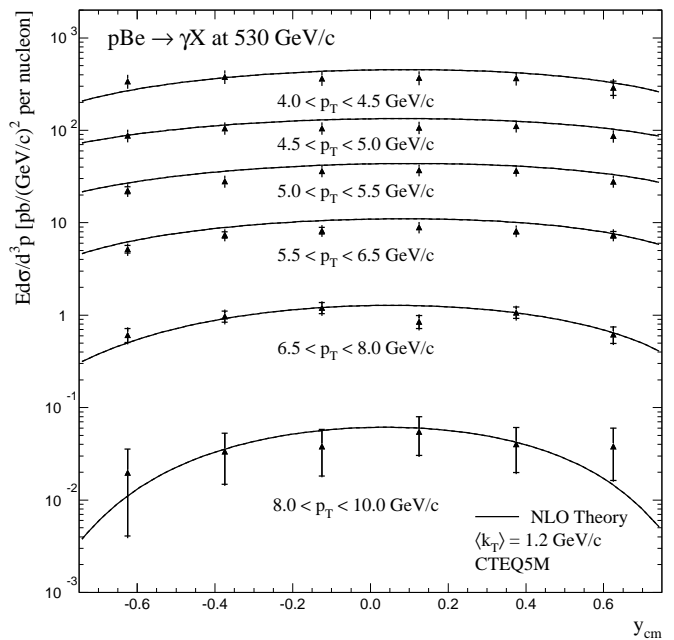}
\caption{
Invariant differential cross sections per nucleon for direct-photon
production as functions of \ycm\ in $p$Be interactions at~530~GeV/$c$
for several intervals in $p_T$. 
The error bars represent the statistical and systematic uncertainties
combined in quadrature; the innermost interval 
indicates the statistical uncertainties.
Overlaid on the data are
$k_T$-enhanced NLO PQCD calculations using the CTEQ5M PDF.}
\label{fig:xs_pt_rap_530_be}
\end{figure}

\begin{figure}
\epsfxsize=\figsize
\epsfbox[\bbcoord]{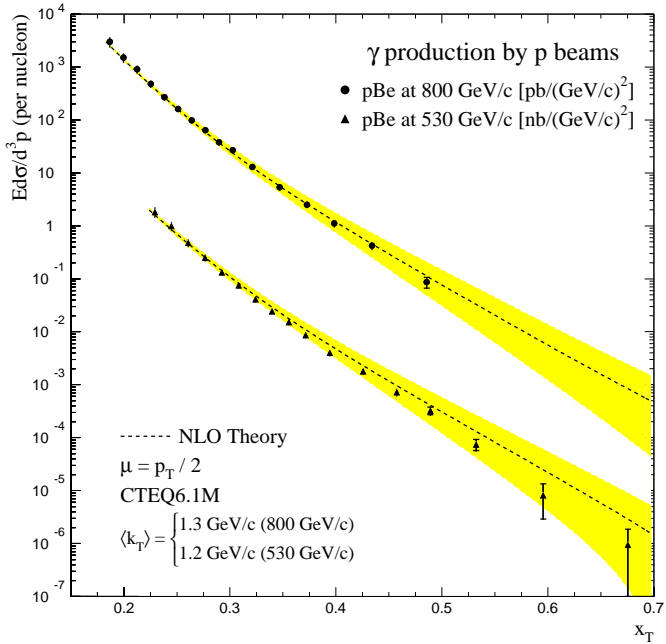}
\caption{
Invariant differential cross sections per nucleon for direct-photon
production as functions of $x_T$ in $p$Be interactions at 800 and
530~GeV/$c$.  
The error bars represent the statistical and systematic uncertainties
combined in quadrature; the innermost interval 
indicates the statistical uncertainties.
Overlaid on the data are $k_T$-enhanced NLO PQCD
calculations.  The shaded region represents the uncertainty band associated
with the CTEQ6.1M PDF set.}
\label{fig:xs_pt_800}
\end{figure}

\begin{figure}
\epsfxsize=\figsize
\epsfbox[\bbcoord]{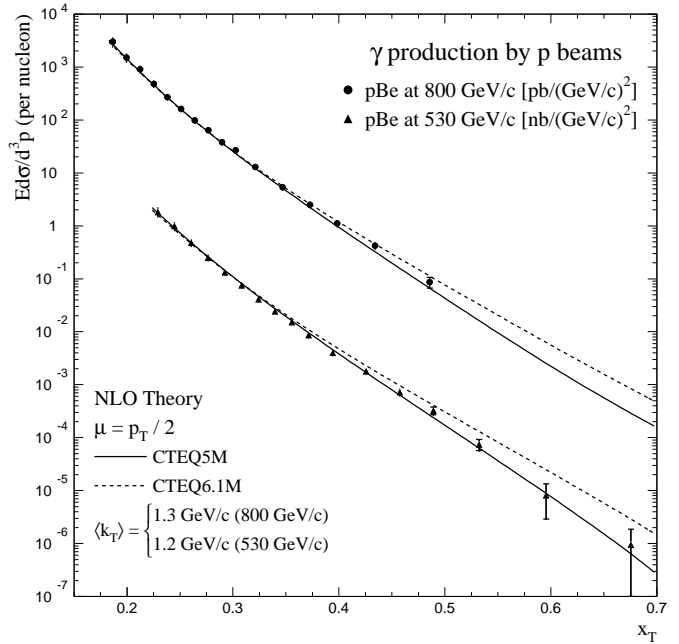}
\caption{
Invariant differential cross sections per nucleon for direct-photon
production as functions of $x_T$ in $p$Be interactions at~800 and
530~GeV/$c$.  
The error bars represent the statistical and systematic uncertainties
combined in quadrature; the innermost interval 
indicates the statistical uncertainties.
Overlaid on the data are $k_T$-enhanced NLO PQCD calculations 
using CTEQ5M and CTEQ6.1M PDF.}
\label{fig:530_800_pt}
\end{figure}

\begin{figure}
\epsfxsize=\figsize
\epsfbox[\bbcoord]{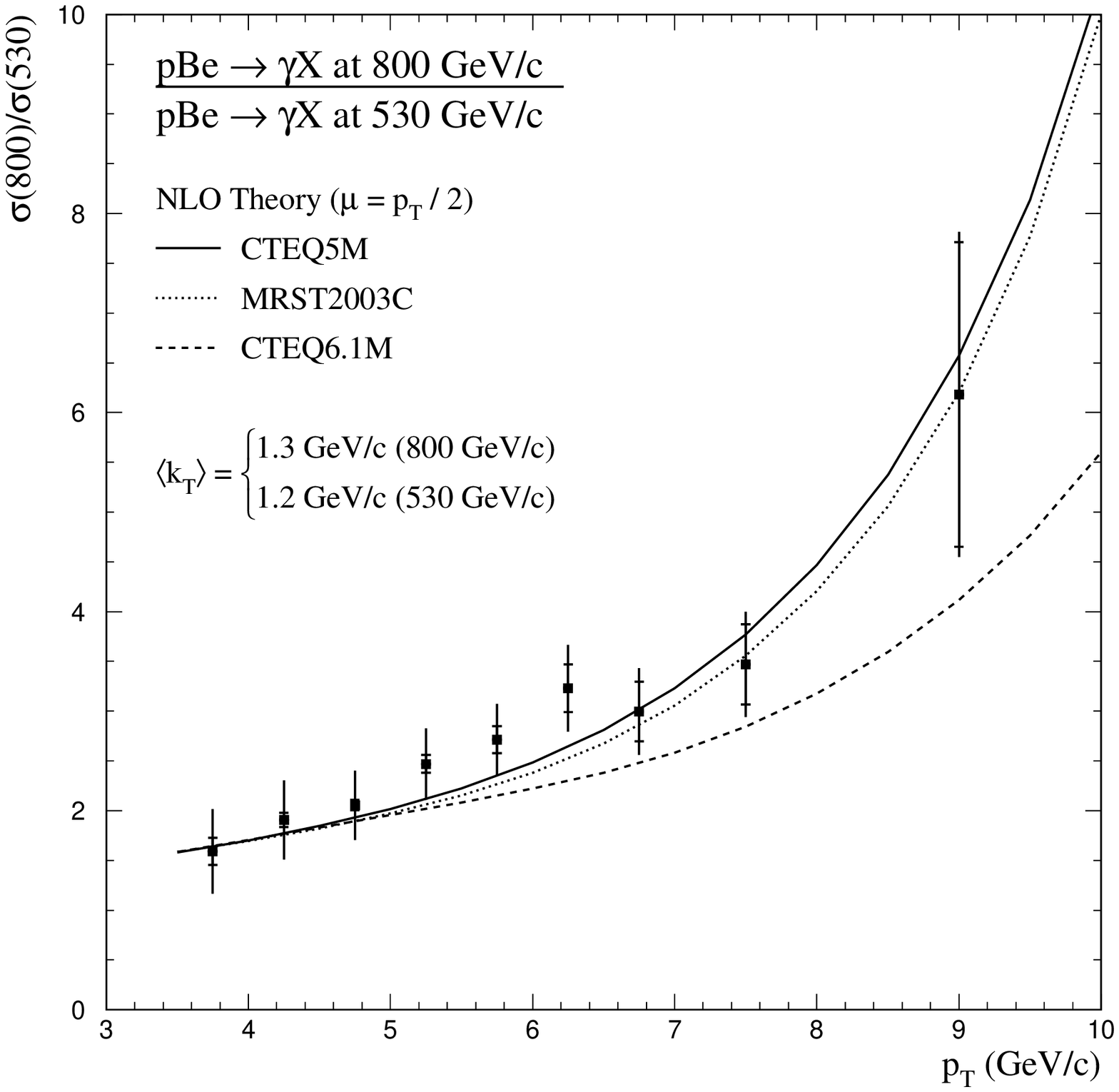}
\caption{The ratio of 
800 to 530~GeV/$c$ proton-beam direct-photon cross sections as a
function of $p_T$. 
The error bars represent the statistical and systematic uncertainties
combined in quadrature; the innermost interval 
indicates the statistical uncertainties.
Results from $k_T$-enhanced NLO PQCD calculations
are overlaid on the data.}
\label{fig:eng_dep_pt}
\end{figure}

\subsection{Comparisons with PQCD}

Next-to-leading order (NLO) PQCD 
calculations~\cite{NLO-aurenche,NLO-berger} using the GRV92
PDF~\cite{grv92} are compared to our measured direct-photon cross
sections for incident $\pi^-$ beam
in Fig.~\ref{fig:xs_pt_515}.  Theoretical results are
presented for three choices of factorization scale: $\mu$ = $p_T$,
$p_T/2$, and $p_T/3$.  The renormalization scale was set equal to the
factorization scale.  Expectations for these choices of $\mu$ lie
significantly below the data, in addition to exhibiting a large
dependence on choice of scale.  Calculations that include higher-order
effects, such as threshold resummation and certain NNLO
diagrams~\cite{laenen,nason,kidonakisowens,sterman-threshold,kidonakisowensNNLO}
significantly reduce scale dependence relative to NLO
calculations, and yield results comparable to the NLO calculation with
$\mu=p_T/2$.  

There is additional theoretical uncertainty associated with the choice
of PDF. Recently published PDF provide individual fits for variations
in fit parameters.  Uncertainty bands based upon these fits are
presented in Fig.~\ref{fig:xs_pt_530_800} for the MRST2001E
PDF set~\cite{mrst2001,mrst2001E} compared with the $pp$ data.  
We conclude there is
insufficient flexibility in the PDF to accommodate the difference
between theory and data.

Not surprisingly, NLO PQCD has reduced scale dependence 
and significantly improved
agreement with data when compared to ratios of direct-photon
cross sections.  The ratio of the 800 to 530~GeV/$c$ direct-photon
cross sections is presented as a function of $x_T=2p_T/\sqrt{s}$ in
Fig.~\ref{fig:eng_dep}.  The ratio between 515~GeV/$c$ $\pi^-$ beam
and 800~GeV/$c$ proton-beam direct-photon cross sections as a function
of~$p_T$ is shown in Fig.~\ref{fig:beam_dep}.  The broad bands on 
the NLO theory in both figures
are bounded by $\mu=p_T/3$ and $p_T$; 
the inner band in Fig.~\ref{fig:eng_dep} represents the 
uncertainty attributed to the PDF.  
The dotted curves in both figures show the results of 
leading-order (LO) PQCD calculations for
$\mu=p_T/2$.
We chose $x_T$ as the plotting variable in 
Fig.~\ref{fig:eng_dep} to compensate for the different average
parton-parton collision energies in the numerator and the denominator;
$p_T$ was used as the plotting variable in  
Fig.~\ref{fig:beam_dep} since the average energy per colliding
valence quark is already similar in the numerator and the denominator.
There is relatively little difference between the LO and NLO curves in
Fig.~\ref{fig:eng_dep}, where the beam particle types are the same for
both the numerator and the denominator.  In marked contrast, however, 
in Fig.~\ref{fig:beam_dep}, where the beam particle types are different
in the numerator and the denominator, the LO and NLO calculations 
are very different, and only the NLO result is consistent with
our data for the PDF employed in these calculations.

Our direct-photon cross-section measurements, as
illustrated in Figs.~\ref{fig:xs_pt_515} and 
\ref{fig:xs_pt_530_800}, as well as our $\pi^0$ cross-section 
measurements reported in previous
publications~\cite{E706-kt,E706-pos-pieta,E706-neg-pieta}, are not
satisfactorily described by the available NLO PQCD calculations.
Similar discrepancies have been observed between PQCD calculations and
other measurements of high-$p_T$ direct-photon and $\pi^0$ 
production~\cite{kt-dp,kt-dp-pi0,ZEUS-kt} and are attributed to the
effect of initial-state soft-gluon radiation (see also
Refs.~\cite{Huston,patrick,frenchpiz}).  The importance of including
gluon emission through the resummation formalism has long been recognized,
and the corresponding calculations have been available for some time
for the diphoton process~\cite{RESBOS,fergani}.  However, these
calculations have only recently been considered for inclusive
direct-photon production~\cite{lilai,li,sterman,sterman-prd}.  While
these developments are encouraging (e.g., see Fig.~2 of
Ref.~\cite{sterman}), the calculations are still incomplete.  In the
absence of a rigorous treatment, the phenomenological
prescription described in Refs.~\cite{kt-dp,E706-kt} 
is used in the following comparisons of 
PQCD with our data.  In this approach, soft-gluon radiation is
parameterized in terms of an effective $\langle k_T\rangle$ that
provides an additional transverse impulse to the incoming partons.
Since the hard-scattering cross sections fall steeply as functions of $p_T$,
the net effect of this additional transverse impulse
is to enhance high-$p_T$ yields.
The $k_T$ enhancement factors (as functions of $p_T$) are evaluated
by computing ratios of results from LO PQCD calculations~\cite{owens} with 
and without $k_T$.  These $k_T$ enhancement factors are applied
to the results of the NLO PQCD calculations.
The $\langle k_T\rangle$ values used in generating the 
$k_T$ enhancement factors are representative of values that 
describe kinematic distributions observed in 
production of high-mass $\pi^0\pi^0$, $\gamma\pi^0$, and 
$\gamma\gamma$ systems~\cite{E706-kt,begel}. 
The $k_T$-enhanced NLO calculations are compared to our
direct-photon measurements in
Fig.~\ref{fig:xs_pt_all_kt}, and provide a much improved
description of our data (relative to the NLO calculations).
Comparisons of the $k_T$-enhanced NLO calculations
to our data as functions of \ycm\ for 
several $p_T$ intervals 
are shown in
Fig.~\ref{fig:xs_pt_rap_530_be} for 530~GeV/$c$
$p$ beam on beryllium.

Results of $k_T$ enhanced NLO PQCD calculations using 
CTEQ6.1M PDF~\cite{cteq6,cteq61}  
are compared to our direct-photon measurements 
for incident protons
in Fig.~\ref{fig:xs_pt_800}.  The CTEQ6.1
PDF uncertainty bands are also shown in that figure.
While the uncertainties
associated with the PDF are small at low $x_T$, they are significantly
larger than the uncertainties in data at moderate and high $x_T$, and are
dominated by the uncertainty in the gluon distribution.  There is also
a large variation in the central value of several PDF sets, as
illustrated in Fig.~\ref{fig:530_800_pt}.  The major difference
between the CTEQ5M~\cite{cteq5} and CTEQ6.1M PDF is that CTEQ6.1M has
a much harder gluon (motivated by measurements of the
inclusive jet cross section at the Tevatron collider).
Cross sections calculated using 
MRST2003C PDF~\cite{mrst2003} are similar to
results using CTEQ5M PDF. 
Figure~\ref{fig:eng_dep_pt} shows the
ratio of the direct-photon cross sections from
800~GeV/$c$ and 530~GeV/$c$ $p$ beams on beryllium,
as a function of $p_T$, compared to
results from $k_T$-enhanced NLO calculations using the MRST2003C,
CTEQ5M, and CTEQ6.1M PDF. 
Calculations with PDF that
have softer gluons provide a better description of our data at high $p_T$.

\section{Conclusions}

High-$p_T$ direct-photon production has been measured in interactions
of 515~GeV/$c$ $\pi^-$ and 530~GeV/$c$ and 800~GeV/$c$ protons
with beryllium and hydrogen targets.  The inclusive
direct-photon cross sections were compared with 
NLO PQCD and $k_T$-enhanced NLO
PQCD calculations for several choices of PDF.  
Without $k_T$ enhancement, 
available NLO PQCD calculations do not satisfactorily represent
our data.
The data are described better by $k_T$ enhanced
NLO PQCD calculations using the softer gluons of CTEQ5 and MRST2003
than the harder gluons of CTEQ6.1.  As mentioned in the Introduction, 
direct-photon
measurements provide potential constraints on the gluon, particularly
at large $x$.  This is
especially relevant since recent global fits have used jet data from
Tevatron Run~I to constrain the large-$x$ gluon---thereby possibly
incorporating new physics effects into the PDF~\cite{cteq61}.  We note
that recent studies have suggested potential incompatibilities between
the small-$x$ gluon determined using HERA data and the large-$x$ gluon
required to fit the Tevatron jet data~\cite{stirling}.  Inclusion of
direct-photon data in the global fits would provide additional
constraints on the gluon distribution, independent of the jet data from 
the Tevatron, thereby enhancing the discovery potential of the 
jet data being acquired in Tevatron Run~II.

\acknowledgments

We thank the U.~S. Department of Energy, the National Science
Foundation, including its Office of International Programs, and the
Universities Grants Commission of India, for their support of this
research.  The staff and management of Fermilab are thanked for their
efforts in making available the beam and computing facilities that
made this work possible.  We are pleased to acknowledge the
contributions of our colleagues on Fermilab experiment E672. We
acknowledge the contributions of the following colleagues for their
help in the operation and upgrade of the Meson West spectrometer:
W.~Dickerson and E.~Pothier from Northeastern University;
J.~T.~Anderson,
E.~Barsotti~Jr.,
H.~Koecher,
P.~Madsen,
D.~Petravick,
R.~Tokarek, J.~Tweed, D.~Allspach, J.~Urbin, and the cryo crews
from Fermi National Accelerator Laboratory;
T.~Haelen, C.~Benson, 
L.~Kuntz, and D.~Ruggiero
from the University of Rochester;
the technical staffs of 
Michigan State University and
%
%
Pennsylvania State University for the construction of the straw tubes
and of the University of Pittsburgh for the silicon detectors. We
thank the following commissioning run collaborators for their
invaluable contributions to the hardware and software infrastructure
of the original Meson West spectrometer: G.~Alverson, G.~Ballocchi,
R.~Benson, D.~Berg, D.~Brown, D.~Carey, T.~Chand, C.~Chandlee,
S.~Easo, W.~Faissler, G.~Glass, I.~Kourbanis, A.~Lanaro,
C.~Nelson~Jr., D.~Orris, B.~Rajaram, K.~Ruddick, A.~Sinanidis, and
G.~Wu.  We also thank S.~Catani, J.Ph.~Guillet, B.~Kniehl, J.~Owens,
G.~Sterman, W.~Vogelsang, and X.-N.~Wang for many helpful discussions
and for providing us with their QCD calculations.

\newpage


\begin{thebibliography}{65}
\expandafter\ifx\csname natexlab\endcsname\relax\def\natexlab#1{#1}\fi
\expandafter\ifx\csname bibnamefont\endcsname\relax
  \def\bibnamefont#1{#1}\fi
\expandafter\ifx\csname bibfnamefont\endcsname\relax
  \def\bibfnamefont#1{#1}\fi
\expandafter\ifx\csname citenamefont\endcsname\relax
  \def\citenamefont#1{#1}\fi
\expandafter\ifx\csname url\endcsname\relax
  \def\url#1{\texttt{#1}}\fi
\expandafter\ifx\csname urlprefix\endcsname\relax\def\urlprefix{URL }\fi
\providecommand{\bibinfo}[2]{#2}
\providecommand{\eprint}[2][]{\url{#2}}

\bibitem[{\citenamefont{Geist et~al.}(1990)}]{geist}
\bibinfo{author}{\bibfnamefont{W.~M.} \bibnamefont{Geist}}
  \bibnamefont{et~al.}, \bibinfo{journal}{Phys. Rept.}
  \textbf{\bibinfo{volume}{197}}, \bibinfo{pages}{263} (\bibinfo{year}{1990}).

\bibitem[{\citenamefont{McCubbin}(1981)}]{mccubbin}
\bibinfo{author}{\bibfnamefont{N.~A.} \bibnamefont{McCubbin}},
  \bibinfo{journal}{Rep. Prog. Phys.} \textbf{\bibinfo{volume}{44}},
  \bibinfo{pages}{65} (\bibinfo{year}{1981}).

\bibitem[{\citenamefont{Owens}(1987)}]{owens}
\bibinfo{author}{\bibfnamefont{J.~F.} \bibnamefont{Owens}},
  \bibinfo{journal}{Rev. Mod. Phys.} \textbf{\bibinfo{volume}{59}},
  \bibinfo{pages}{465} (\bibinfo{year}{1987}).

\bibitem[{\citenamefont{T.Ferbel and Molzon}(1984)}]{molzon}
\bibinfo{author}{\bibnamefont{T.Ferbel}} \bibnamefont{and}
  \bibinfo{author}{\bibfnamefont{W.~R.} \bibnamefont{Molzon}},
  \bibinfo{journal}{Rev. Mod. Phys.} \textbf{\bibinfo{volume}{56}},
  \bibinfo{pages}{181} (\bibinfo{year}{1984}).

\bibitem[{\citenamefont{Huston et~al.}(1998)}]{huston-uncertainty}
\bibinfo{author}{\bibfnamefont{J.}~\bibnamefont{Huston}} \bibnamefont{et~al.},
  \bibinfo{journal}{Phys. Rev.} \textbf{\bibinfo{volume}{D58}},
  \bibinfo{pages}{114034} (\bibinfo{year}{1998}).

\bibitem[{\citenamefont{Lai et~al.}(1997)}]{cteq4}
\bibinfo{author}{\bibfnamefont{H.~L.} \bibnamefont{Lai}} \bibnamefont{et~al.},
  \bibinfo{journal}{Phys.~Rev.} \textbf{\bibinfo{volume}{D55}},
  \bibinfo{pages}{1280} (\bibinfo{year}{1997}).

\bibitem[{\citenamefont{Gluck et~al.}(1992)\citenamefont{Gluck, Reya, and
  Vogt}}]{grv92}
\bibinfo{author}{\bibfnamefont{M.}~\bibnamefont{Gluck}},
  \bibinfo{author}{\bibfnamefont{E.}~\bibnamefont{Reya}}, \bibnamefont{and}
  \bibinfo{author}{\bibfnamefont{A.}~\bibnamefont{Vogt}}, \bibinfo{journal}{Z.
  Phys.} \textbf{\bibinfo{volume}{C53}}, \bibinfo{pages}{127}
  (\bibinfo{year}{1992}).

\bibitem[{\citenamefont{Martin et~al.}(1998)\citenamefont{Martin, Roberts,
  Stirling, and Thorne}}]{mrst98}
\bibinfo{author}{\bibfnamefont{A.~D.} \bibnamefont{Martin}},
  \bibinfo{author}{\bibfnamefont{R.~G.} \bibnamefont{Roberts}},
  \bibinfo{author}{\bibfnamefont{W.~J.} \bibnamefont{Stirling}},
  \bibnamefont{and} \bibinfo{author}{\bibfnamefont{R.~S.}
  \bibnamefont{Thorne}}, \bibinfo{journal}{Eur. Phys. J.}
  \textbf{\bibinfo{volume}{C4}}, \bibinfo{pages}{463} (\bibinfo{year}{1998}).

\bibitem[{\citenamefont{Apanasevich et~al.}(1998{\natexlab{a}})}]{E706-kt}
\bibinfo{author}{\bibfnamefont{L.}~\bibnamefont{Apanasevich}}
  \bibnamefont{et~al.}, \bibinfo{journal}{Phys. Rev. Lett.}
  \textbf{\bibinfo{volume}{81}}, \bibinfo{pages}{2642}
  (\bibinfo{year}{1998}{\natexlab{a}}).

\bibitem[{E70({\natexlab{a}})}]{E706-pos-pieta}
\bibinfo{note}{L. Apanasevich et al., Phys. Rev. {\bf D68}, 052001 (2003). A
  computational error in the correction of measured cross sections for
  contributions from minority particles in secondary beams affected the
  reported cross sections for $\pi^0$ and $\eta$ production by 530~GeV/$c$
  protons. As a result, the cross sections listed in the tables for 530~GeV/$c$
  protons in the publication referenced here should be increased by 3.4$\%$, a
  change that falls well within the quoted systematic uncertainties.}

\bibitem[{\citenamefont{Apanasevich et~al.}(2004)}]{E706-neg-pieta}
\bibinfo{author}{\bibfnamefont{L.}~\bibnamefont{Apanasevich}}
  \bibnamefont{et~al.}, \bibinfo{journal}{Phys. Rev.}
  \textbf{\bibinfo{volume}{D69}}, \bibinfo{pages}{032003}
  (\bibinfo{year}{2004}).

\bibitem[{\citenamefont{Apanasevich et~al.}(1997)}]{E706-charm}
\bibinfo{author}{\bibfnamefont{L.}~\bibnamefont{Apanasevich}}
  \bibnamefont{et~al.}, \bibinfo{journal}{Phys. Rev.}
  \textbf{\bibinfo{volume}{D56}}, \bibinfo{pages}{1391} (\bibinfo{year}{1997}).

\bibitem[{\citenamefont{Apanasevich et~al.}(2000)}]{E706-omega}
\bibinfo{author}{\bibfnamefont{L.}~\bibnamefont{Apanasevich}}
  \bibnamefont{et~al.} (\bibinfo{year}{2000}), \eprint{hep-ex/0004012}.

\bibitem[{\citenamefont{Alverson et~al.}(1993)}]{E706-88}
\bibinfo{author}{\bibfnamefont{G.}~\bibnamefont{Alverson}}
  \bibnamefont{et~al.}, \bibinfo{journal}{Phys. Rev.}
  \textbf{\bibinfo{volume}{D48}}, \bibinfo{pages}{5} (\bibinfo{year}{1993}).

\bibitem[{\citenamefont{Alverson et~al.}(1994)}]{E706-88-recoil}
\bibinfo{author}{\bibfnamefont{G.}~\bibnamefont{Alverson}}
  \bibnamefont{et~al.}, \bibinfo{journal}{Phys. Rev.}
  \textbf{\bibinfo{volume}{D49}}, \bibinfo{pages}{3106} (\bibinfo{year}{1994}).

\bibitem[{not({\natexlab{a}})}]{note-88}
\bibinfo{note}{The event samples presented in this paper exclude data collected
  during the 1987-1988 commissioning run of E706 (presented in
  Ref.~\cite{E706-88} and references therein). Although our earlier data are
  generally consistent with the results reported here, once systematic
  differences between the two analyses are properly accounted for, the results
  from the initial run are superseded by our current higher statistics, more
  precise data samples. Theoretical calculations presented here also supersede
  earlier calculations.}

\bibitem[{\citenamefont{Allspach et~al.}(1991)}]{h2target}
\bibinfo{author}{\bibfnamefont{D.}~\bibnamefont{Allspach}}
  \bibnamefont{et~al.}, \bibinfo{journal}{Adv. Cryog. Eng.}
  \textbf{\bibinfo{volume}{37}}, \bibinfo{pages}{1495} (\bibinfo{year}{1991}).

\bibitem[{\citenamefont{Bromberg et~al.}(1991)}]{E706-STDC}
\bibinfo{author}{\bibfnamefont{C.}~\bibnamefont{Bromberg}}
  \bibnamefont{et~al.}, \bibinfo{journal}{Nucl. Inst. \& Meth.}
  \textbf{\bibinfo{volume}{A307}}, \bibinfo{pages}{292} (\bibinfo{year}{1991}).

\bibitem[{\citenamefont{Apanasevich
  et~al.}(1998{\natexlab{b}})}]{E706-calibration}
\bibinfo{author}{\bibfnamefont{L.}~\bibnamefont{Apanasevich}}
  \bibnamefont{et~al.}, \bibinfo{journal}{Nucl. Inst. \& Meth.}
  \textbf{\bibinfo{volume}{A417}}, \bibinfo{pages}{50}
  (\bibinfo{year}{1998}{\natexlab{b}}).

\bibitem[{\citenamefont{Gribushin et~al.}(1996)}]{psi90}
\bibinfo{author}{\bibfnamefont{A.}~\bibnamefont{Gribushin}}
  \bibnamefont{et~al.}, \bibinfo{journal}{Phys. Rev.}
  \textbf{\bibinfo{volume}{D53}}, \bibinfo{pages}{4723} (\bibinfo{year}{1996}).

\bibitem[{\citenamefont{Gribushin et~al.}(2000)}]{psi91}
\bibinfo{author}{\bibfnamefont{A.}~\bibnamefont{Gribushin}}
  \bibnamefont{et~al.}, \bibinfo{journal}{Phys. Rev.}
  \textbf{\bibinfo{volume}{D62}}, \bibinfo{pages}{012001}
  (\bibinfo{year}{2000}).

\bibitem[{\citenamefont{Koreshev et~al.}(1996)}]{chi90}
\bibinfo{author}{\bibfnamefont{V.}~\bibnamefont{Koreshev}}
  \bibnamefont{et~al.}, \bibinfo{journal}{Phys. Rev. Lett.}
  \textbf{\bibinfo{volume}{77}}, \bibinfo{pages}{4294} (\bibinfo{year}{1996}).

\bibitem[{\citenamefont{Jesik et~al.}(1995)}]{B90}
\bibinfo{author}{\bibfnamefont{R.}~\bibnamefont{Jesik}} \bibnamefont{et~al.},
  \bibinfo{journal}{Phys. Rev. Lett.} \textbf{\bibinfo{volume}{74}},
  \bibinfo{pages}{495} (\bibinfo{year}{1995}).

\bibitem[{\citenamefont{Striley}(1996)}]{striley}
\bibinfo{author}{\bibfnamefont{D.}~\bibnamefont{Striley}}, Ph.D. thesis,
  \bibinfo{school}{University of Missouri-Columbia} (\bibinfo{year}{1996}).

\bibitem[{E70({\natexlab{b}})}]{E706-trigger}
\bibinfo{note}{L.~Sorrell, {\it The E706 Trigger System}, E706 Note 201, 1994
  (unpublished); Ph.D. thesis, Michigan State University, 1995; G.~Osborne,
  Ph.D. thesis, University of Rochester, 1996.}

\bibitem[{BH()}]{BH}
\bibinfo{note}{During the 1990 run, the beam hole definition was implemented
  using a single scintillation counter. An array of four scintillation counters
  was used for the beam hole definition during the 1991-1992 run.}

\bibitem[{her()}]{herwig61}
\bibinfo{note}{G.~Corcella et al., JHEP, {\bf 01}, (2001) 010, {\sc
  herwig}~v6.1. {\sc herwig}~v6.1 was used as the event generator for the
  vertex efficiency studies. The relative normalizations of the simulated
  samples from various targets were tuned to match the data.}

\bibitem[{gea()}]{geant}
\bibinfo{note}{F. Carminati et al., GEANT: Detector Description and Simulation
  Tool, {\rm CERN Program Library Long Writeup W5013}, 1993.}

\bibitem[{\citenamefont{Marchesini et~al.}(1992)}]{herwig56}
\bibinfo{author}{\bibfnamefont{G.}~\bibnamefont{Marchesini}}
  \bibnamefont{et~al.}, \bibinfo{journal}{Comput.~Phys.~Commun.}
  \textbf{\bibinfo{volume}{67}}, \bibinfo{pages}{465} (\bibinfo{year}{1992}),
  \bibinfo{note}{{\sc herwig}~v5.6}.

\bibitem[{apa()}]{apana}
\bibinfo{note}{L.~Apanasevich, Ph.D. thesis, Michigan State University (in
  preparation)}.

\bibitem[{not({\natexlab{b}})}]{note-etatopi}
\bibinfo{note}{Our measured $\eta/\pi^0$ ratios as functions of $p_T$ were fit
  to constants for $p_T~>$~3.0~GeV/$c$. For the 515~GeV/$c$ $\pi^-$Be
  interactions, the resulting $\eta/\pi^0$ is 0.48$\pm$0.01(stat). The relevant
  data is shown in Fig.~15 of Ref.~\cite{E706-neg-pieta}. The corresponding
  $\eta/\pi^0$ from the 800~GeV/$c$ $p$Be interactions is 0.42$\pm$0.01(stat),
  while for the 530~GeV/$c$ $p$Be interactions the result is
  0.45$\pm$0.01(stat)~\cite{E706-pos-pieta}. The generated $\eta$ spectra were
  weighted to match the shape of our measured $\pi^0$ spectra, invoking
  appropriate relative normalization factors.}

\bibitem[{not({\natexlab{c}})}]{note-herwig-omega}
\bibinfo{note}{As illustrated in Fig.~10.1 of Ref~\cite{debarbaro}, {\sc
  herwig}~v5.6 provides a good description of the $\omega$ to $\pi^0$ ratio
  observed in our data. However, Ref.~\cite{E706-omega} shows that {\sc
  herwig}~v6.1 generates a much smaller $\omega$ to $\pi^0$ production ratio.
  {\sc herwig}~v5.6 was used to generate the DGS samples employed in the
  evaluation of the direct-photon backgrounds.}

\bibitem[{not({\natexlab{d}})}]{note-etaprime}
\bibinfo{note}{The $\eta^\prime/\pi^0$ ratio from {\sc herwig}~v5.6 is
  $\approx0.2$. A measurement of this high-$p_T$ production ratio yields
  0.85$\pm$0.24 (after updating the $\eta^\prime\rightarrow\gamma\gamma$
  branching ratio)~\cite{etaprimeratio,pdg}. This difference has been included
  as a contribution to the systematic uncertainties in the direct-photon
  background.}

\bibitem[{\citenamefont{Begel}(1999)}]{begel}
\bibinfo{author}{\bibfnamefont{M.}~\bibnamefont{Begel}}, Ph.D. thesis,
  \bibinfo{school}{University of Rochester} (\bibinfo{year}{1999}).

\bibitem[{\citenamefont{Diakonou et~al.}(1980)}]{etaprimeratio}
\bibinfo{author}{\bibfnamefont{M.}~\bibnamefont{Diakonou}}
  \bibnamefont{et~al.}, \bibinfo{journal}{Phys. Lett.}
  \textbf{\bibinfo{volume}{89B}}, \bibinfo{pages}{432} (\bibinfo{year}{1980}).

\bibitem[{\citenamefont{Groom et~al.}(2000)}]{pdg}
\bibinfo{author}{\bibfnamefont{D.~E.} \bibnamefont{Groom}}
  \bibnamefont{et~al.}, \bibinfo{journal}{Eur. Phys. J.}
  \textbf{\bibinfo{volume}{C15}}, \bibinfo{pages}{1} (\bibinfo{year}{2000}).

\bibitem[{\citenamefont{Aurenche et~al.}(1984)}]{NLO-aurenche}
\bibinfo{author}{\bibfnamefont{P.}~\bibnamefont{Aurenche}}
  \bibnamefont{et~al.}, \bibinfo{journal}{Phys. Lett.}
  \textbf{\bibinfo{volume}{B140}}, \bibinfo{pages}{87} (\bibinfo{year}{1984}).

\bibitem[{\citenamefont{Berger and Qiu}(1991)}]{NLO-berger}
\bibinfo{author}{\bibfnamefont{E.~L.} \bibnamefont{Berger}} \bibnamefont{and}
  \bibinfo{author}{\bibfnamefont{J.}~\bibnamefont{Qiu}},
  \bibinfo{journal}{Phys. Rev.} \textbf{\bibinfo{volume}{D44}},
  \bibinfo{pages}{2002} (\bibinfo{year}{1991}).

\bibitem[{\citenamefont{Lafferty and Wyatt}(1995)}]{laff}
\bibinfo{author}{\bibfnamefont{G.~D.} \bibnamefont{Lafferty}} \bibnamefont{and}
  \bibinfo{author}{\bibfnamefont{T.~R.} \bibnamefont{Wyatt}},
  \bibinfo{journal}{Nucl. Instrum. Methods Phys. Res.}
  \textbf{\bibinfo{volume}{A355}}, \bibinfo{pages}{541} (\bibinfo{year}{1995}).

\bibitem[{wan()}]{wang-note}
\bibinfo{note}{Xin-Nian Wang, Nucl. Phys. {\bf A661}, 609 (1999), and private
  communication with Xin-Nian Wang regarding the use of this program at our
  beam energies.}

\bibitem[{\citenamefont{Laenen et~al.}(1998)\citenamefont{Laenen, Oderda, and
  Sterman}}]{laenen}
\bibinfo{author}{\bibfnamefont{E.}~\bibnamefont{Laenen}},
  \bibinfo{author}{\bibfnamefont{G.}~\bibnamefont{Oderda}}, \bibnamefont{and}
  \bibinfo{author}{\bibfnamefont{G.}~\bibnamefont{Sterman}},
  \bibinfo{journal}{Phys. Lett.} \textbf{\bibinfo{volume}{B438}},
  \bibinfo{pages}{173} (\bibinfo{year}{1998}).

\bibitem[{\citenamefont{Catani et~al.}(1999)}]{nason}
\bibinfo{author}{\bibfnamefont{S.}~\bibnamefont{Catani}} \bibnamefont{et~al.},
  \bibinfo{journal}{JHEP} \textbf{\bibinfo{volume}{03}}, \bibinfo{pages}{025}
  (\bibinfo{year}{1999}).

\bibitem[{\citenamefont{Kidonakis and Owens}(2000)}]{kidonakisowens}
\bibinfo{author}{\bibfnamefont{N.}~\bibnamefont{Kidonakis}} \bibnamefont{and}
  \bibinfo{author}{\bibfnamefont{J.~F.} \bibnamefont{Owens}},
  \bibinfo{journal}{Phys. Rev.} \textbf{\bibinfo{volume}{D61}},
  \bibinfo{pages}{094004} (\bibinfo{year}{2000}).

\bibitem[{\citenamefont{Sterman and Vogelsang}(2001)}]{sterman-threshold}
\bibinfo{author}{\bibfnamefont{G.}~\bibnamefont{Sterman}} \bibnamefont{and}
  \bibinfo{author}{\bibfnamefont{W.}~\bibnamefont{Vogelsang}},
  \bibinfo{journal}{JHEP} \textbf{\bibinfo{volume}{02}}, \bibinfo{pages}{016}
  (\bibinfo{year}{2001}).

\bibitem[{\citenamefont{Kidonakis and Owens}(2003)}]{kidonakisowensNNLO}
\bibinfo{author}{\bibfnamefont{N.}~\bibnamefont{Kidonakis}} \bibnamefont{and}
  \bibinfo{author}{\bibfnamefont{J.~F.} \bibnamefont{Owens}}
  (\bibinfo{year}{2003}), \eprint{hep-ph/0307352}.

\bibitem[{\citenamefont{Martin et~al.}(2002)\citenamefont{Martin, Roberts,
  Stirling, and Thorne}}]{mrst2001}
\bibinfo{author}{\bibfnamefont{A.~D.} \bibnamefont{Martin}},
  \bibinfo{author}{\bibfnamefont{R.~G.} \bibnamefont{Roberts}},
  \bibinfo{author}{\bibfnamefont{W.~J.} \bibnamefont{Stirling}},
  \bibnamefont{and} \bibinfo{author}{\bibfnamefont{R.~S.}
  \bibnamefont{Thorne}}, \bibinfo{journal}{Eur. Phys. J.}
  \textbf{\bibinfo{volume}{C23}}, \bibinfo{pages}{73} (\bibinfo{year}{2002}).

\bibitem[{\citenamefont{Martin et~al.}(2003{\natexlab{a}})\citenamefont{Martin,
  Roberts, Stirling, and Thorne}}]{mrst2001E}
\bibinfo{author}{\bibfnamefont{A.~D.} \bibnamefont{Martin}},
  \bibinfo{author}{\bibfnamefont{R.~G.} \bibnamefont{Roberts}},
  \bibinfo{author}{\bibfnamefont{W.~J.} \bibnamefont{Stirling}},
  \bibnamefont{and} \bibinfo{author}{\bibfnamefont{R.~S.}
  \bibnamefont{Thorne}}, \bibinfo{journal}{Eur. Phys. J.}
  \textbf{\bibinfo{volume}{C28}}, \bibinfo{pages}{455}
  (\bibinfo{year}{2003}{\natexlab{a}}).

\bibitem[{\citenamefont{Apanasevich et~al.}(1999)}]{kt-dp}
\bibinfo{author}{\bibfnamefont{L.}~\bibnamefont{Apanasevich}}
  \bibnamefont{et~al.}, \bibinfo{journal}{Phys. Rev.}
  \textbf{\bibinfo{volume}{D59}}, \bibinfo{pages}{074007}
  (\bibinfo{year}{1999}).

\bibitem[{\citenamefont{Apanasevich et~al.}(2001)}]{kt-dp-pi0}
\bibinfo{author}{\bibfnamefont{L.}~\bibnamefont{Apanasevich}}
  \bibnamefont{et~al.}, \bibinfo{journal}{Phys. Rev.}
  \textbf{\bibinfo{volume}{D63}}, \bibinfo{pages}{014009}
  (\bibinfo{year}{2001}).

\bibitem[{\citenamefont{Chekanov et~al.}(2001)}]{ZEUS-kt}
\bibinfo{author}{\bibfnamefont{S.}~\bibnamefont{Chekanov}}
  \bibnamefont{et~al.}, \bibinfo{journal}{Phys. Lett.}
  \textbf{\bibinfo{volume}{511B}}, \bibinfo{pages}{19} (\bibinfo{year}{2001}).

\bibitem[{\citenamefont{Huston et~al.}(1995)}]{Huston}
\bibinfo{author}{\bibfnamefont{J.}~\bibnamefont{Huston}} \bibnamefont{et~al.},
  \bibinfo{journal}{Phys. Rev.} \textbf{\bibinfo{volume}{D51}},
  \bibinfo{pages}{6139} (\bibinfo{year}{1995}).

\bibitem[{\citenamefont{Aurenche et~al.}(1999)}]{patrick}
\bibinfo{author}{\bibfnamefont{P.}~\bibnamefont{Aurenche}}
  \bibnamefont{et~al.}, \bibinfo{journal}{Eur. Phys. J.}
  \textbf{\bibinfo{volume}{C9}}, \bibinfo{pages}{107} (\bibinfo{year}{1999}).

\bibitem[{\citenamefont{Aurenche et~al.}(2000)}]{frenchpiz}
\bibinfo{author}{\bibfnamefont{P.}~\bibnamefont{Aurenche}}
  \bibnamefont{et~al.}, \bibinfo{journal}{Eur. Phys. J.}
  \textbf{\bibinfo{volume}{C13}}, \bibinfo{pages}{347} (\bibinfo{year}{2000}).

\bibitem[{\citenamefont{Bal\'{a}zs et~al.}(1998)}]{RESBOS}
\bibinfo{author}{\bibfnamefont{C.}~\bibnamefont{Bal\'{a}zs}}
  \bibnamefont{et~al.}, \bibinfo{journal}{Phys. Rev.}
  \textbf{\bibinfo{volume}{D57}}, \bibinfo{pages}{6934} (\bibinfo{year}{1998}).

\bibitem[{\citenamefont{Chiappetta et~al.}(1995)\citenamefont{Chiappetta,
  Fergani, and Guillet}}]{fergani}
\bibinfo{author}{\bibfnamefont{P.}~\bibnamefont{Chiappetta}},
  \bibinfo{author}{\bibfnamefont{R.}~\bibnamefont{Fergani}}, \bibnamefont{and}
  \bibinfo{author}{\bibfnamefont{J.}~\bibnamefont{Guillet}},
  \bibinfo{journal}{Phys. Lett.} \textbf{\bibinfo{volume}{B348}},
  \bibinfo{pages}{646} (\bibinfo{year}{1995}).

\bibitem[{\citenamefont{Lai and Li}(1998)}]{lilai}
\bibinfo{author}{\bibfnamefont{H.-L.} \bibnamefont{Lai}} \bibnamefont{and}
  \bibinfo{author}{\bibfnamefont{H.}~\bibnamefont{Li}}, \bibinfo{journal}{Phys.
  Rev.} \textbf{\bibinfo{volume}{D58}}, \bibinfo{pages}{114020}
  (\bibinfo{year}{1998}).

\bibitem[{\citenamefont{Li}(1999)}]{li}
\bibinfo{author}{\bibfnamefont{H.}~\bibnamefont{Li}}, \bibinfo{journal}{Phys.
  Lett.} \textbf{\bibinfo{volume}{B454}}, \bibinfo{pages}{328}
  (\bibinfo{year}{1999}).

\bibitem[{\citenamefont{Laenen et~al.}(2000)\citenamefont{Laenen, Sterman, and
  Vogelsang}}]{sterman}
\bibinfo{author}{\bibfnamefont{E.}~\bibnamefont{Laenen}},
  \bibinfo{author}{\bibfnamefont{G.}~\bibnamefont{Sterman}}, \bibnamefont{and}
  \bibinfo{author}{\bibfnamefont{W.}~\bibnamefont{Vogelsang}},
  \bibinfo{journal}{Phys. Rev. Lett.} \textbf{\bibinfo{volume}{84}},
  \bibinfo{pages}{4296} (\bibinfo{year}{2000}).

\bibitem[{\citenamefont{Laenen et~al.}(2001)\citenamefont{Laenen, Sterman, and
  Vogelsang}}]{sterman-prd}
\bibinfo{author}{\bibfnamefont{E.}~\bibnamefont{Laenen}},
  \bibinfo{author}{\bibfnamefont{G.}~\bibnamefont{Sterman}}, \bibnamefont{and}
  \bibinfo{author}{\bibfnamefont{W.}~\bibnamefont{Vogelsang}},
  \bibinfo{journal}{Phys. Rev.} \textbf{\bibinfo{volume}{D63}},
  \bibinfo{pages}{114018} (\bibinfo{year}{2001}).

\bibitem[{\citenamefont{Pumplin et~al.}(2002)}]{cteq6}
\bibinfo{author}{\bibfnamefont{J.}~\bibnamefont{Pumplin}} \bibnamefont{et~al.},
  \bibinfo{journal}{JHEP} \textbf{\bibinfo{volume}{07}}, \bibinfo{pages}{012}
  (\bibinfo{year}{2002}).

\bibitem[{\citenamefont{Stump et~al.}(2003)}]{cteq61}
\bibinfo{author}{\bibfnamefont{D.}~\bibnamefont{Stump}} \bibnamefont{et~al.},
  \bibinfo{journal}{JHEP} \textbf{\bibinfo{volume}{10}}, \bibinfo{pages}{046}
  (\bibinfo{year}{2003}).

\bibitem[{\citenamefont{Lai et~al.}(2000)}]{cteq5}
\bibinfo{author}{\bibfnamefont{H.~L.} \bibnamefont{Lai}} \bibnamefont{et~al.},
  \bibinfo{journal}{Eur. Phys. J.} \textbf{\bibinfo{volume}{C12}},
  \bibinfo{pages}{375} (\bibinfo{year}{2000}).

\bibitem[{\citenamefont{Martin et~al.}(2003{\natexlab{b}})\citenamefont{Martin,
  Roberts, Stirling, and Thorne}}]{mrst2003}
\bibinfo{author}{\bibfnamefont{A.~D.} \bibnamefont{Martin}},
  \bibinfo{author}{\bibfnamefont{R.~G.} \bibnamefont{Roberts}},
  \bibinfo{author}{\bibfnamefont{W.~J.} \bibnamefont{Stirling}},
  \bibnamefont{and} \bibinfo{author}{\bibfnamefont{R.~S.} \bibnamefont{Thorne}}
  (\bibinfo{year}{2003}{\natexlab{b}}), \eprint{hep-ph/0307262}.

\bibitem[{sti()}]{stirling}
\bibinfo{note}{A.~D.~Martin, R.~G.~Roberts, W.~J.~Stirling, and R.~S.~Thorne,
  (2003), hep-ph/0308087; J. Stirling, presentation at the Collider Physics
  conference, Kavli Institute for Theoretical Physics, Santa Barbara, January
  2004.}

\bibitem[{\citenamefont{de~Barbaro}(1995)}]{debarbaro}
\bibinfo{author}{\bibfnamefont{L.}~\bibnamefont{de~Barbaro}}, Ph.D. thesis,
  \bibinfo{school}{University of Rochester} (\bibinfo{year}{1995}).

\end{thebibliography}

\vfil

\newpage
\clearpage

\onecolumngrid

\appendix*

\section{Tabulated Cross Sections}

\begin{table}[h]
\caption{Invariant differential cross sections $\left( \DIFFXS \right)$ 
per nucleon for direct-photon production in \pBe\ collisions at 800 and
530~GeV/$c$, and \piBe\ collisions at 515~GeV/$c$ as functions of $p_T$.}
\begin{tabular}{r@{ -- }l r@{ }l r@{ }l r@{ }l}
\hline
\hline
\multicolumn{2}{c}{$p_T$}   &\multicolumn{2}{c}{${\mit p}$Be at 800~GeV/$c$}   &\multicolumn{2}{c}{${\mit p}$Be at 530~GeV/$c$}   &\multicolumn{2}{c}{${\mit \pi^{-}}$Be at 515~GeV/$c$} \\
\multicolumn{2}{c}{ }   &\multicolumn{2}{c}{$-1.0\le\ycm\le 0.5$}   &\multicolumn{2}{c}{$-0.75\le\ycm\le 0.75$}   &\multicolumn{2}{c}{$-0.75\le\ycm\le 0.75$} \\
\multicolumn{2}{c}{(GeV/$c$)}    &\multicolumn{2}{c}{$[\rm{nb}$/(GeV/$c)^2]$}    &\multicolumn{2}{c}{$[\rm{nb}$/(GeV/$c)^2]$}    &\multicolumn{2}{c}{$[\rm{nb}$/(GeV/$c)^2]$}  \\
\hline
3.50 & 3.75 & 
 {2.99${\,}\pm$} & {0.24${}\pm{}$0.62} &  {1.83${\,}\pm$} & {0.17${}\pm{}$0.37} &  {1.79${\,}\pm$} & {0.10${}\pm{}$0.33} \\
3.75 & 4.00 & 
 {1.500${\,}\pm$} & {0.091${}\pm{}$0.29} &  {0.993${\,}\pm$} & {0.062${}\pm{}$0.18} &  {0.924${\,}\pm$} & {0.036${}\pm{}$0.16} \\
4.00 & 4.25 & 
 {0.908${\,}\pm$} & {0.031${}\pm{}$0.16} &  {0.478${\,}\pm$} & {0.019${}\pm{}$0.081} &  {0.5058${\,}\pm$} & {0.0081${}\pm{}$0.080} \\
4.25 & 4.50 & 
 {0.482${\,}\pm$} & {0.011${}\pm{}$0.079} &  {0.2508${\,}\pm$} & {0.0047${}\pm{}$0.040} &  {0.2891${\,}\pm$} & {0.0052${}\pm{}$0.043} \\
4.50 & 4.75 & 
 {0.2679${\,}\pm$} & {0.0068${}\pm{}$0.041} &  {0.1334${\,}\pm$} & {0.0030${}\pm{}$0.020} &  {0.1736${\,}\pm$} & {0.0034${}\pm{}$0.024} \\
4.75 & 5.00 & 
 {0.1616${\,}\pm$} & {0.0042${}\pm{}$0.023} &  {0.0757${\,}\pm$} & {0.0021${}\pm{}$0.011} &  {0.1084${\,}\pm$} & {0.0024${}\pm{}$0.014} \\
\hline
\multicolumn{2}{c}{ }    &\multicolumn{2}{c}{$[\rm{pb}$/(GeV/$c)^2]$}    &\multicolumn{2}{c}{$[\rm{pb}$/(GeV/$c)^2]$}    &\multicolumn{2}{c}{$[\rm{pb}$/(GeV/$c)^2]$}  \\
\hline
5.00 & 5.25 & 
 {98.4${\,}\pm$} & {3.1${}\pm{}$14} &  {41.3${\,}\pm$} & {1.4${}\pm{}$5.5} &  {70.8${\,}\pm$} & {1.8${}\pm{}$9.0} \\
5.25 & 5.50 & 
 {63.7${\,}\pm$} & {2.3${}\pm{}$8.4} &  {24.4${\,}\pm$} & {1.0${}\pm{}$3.2} &  {42.7${\,}\pm$} & {1.3${}\pm{}$5.2} \\
5.50 & 5.75 & 
 {37.7${\,}\pm$} & {1.6${}\pm{}$4.8} &  {15.16${\,}\pm$} & {0.74${}\pm{}$1.9} &  {30.6${\,}\pm$} & {1.0${}\pm{}$3.7} \\
5.75 & 6.00 & 
 {26.7${\,}\pm$} & {1.3${}\pm{}$3.3} &  {8.60${\,}\pm$} & {0.55${}\pm{}$1.1} &  {20.02${\,}\pm$} & {0.77${}\pm{}$2.4} \\
6.00 & 6.50 & 
 {12.86${\,}\pm$} & {0.52${}\pm{}$1.6} &  {3.98${\,}\pm$} & {0.25${}\pm{}$0.49} &  {10.38${\,}\pm$} & {0.37${}\pm{}$1.2} \\
6.50 & 7.00 & 
 {5.36${\,}\pm$} & {0.30${}\pm{}$0.63} &  {1.79${\,}\pm$} & {0.15${}\pm{}$0.22} &  {4.74${\,}\pm$} & {0.23${}\pm{}$0.55} \\
7.00 & 7.50 & 
 {2.51${\,}\pm$} & {0.19${}\pm{}$0.29} &  {0.718${\,}\pm$} & {0.086${}\pm{}$0.089} &  {2.12${\,}\pm$} & {0.15${}\pm{}$0.25} \\
7.50 & 8.00 & 
 {1.11${\,}\pm$} & {0.12${}\pm{}$0.13} &  {0.324${\,}\pm$} & {0.054${}\pm{}$0.041} &  {1.148${\,}\pm$} & {0.099${}\pm{}$0.14} \\
8.00 & 9.00 & 
 {0.420${\,}\pm$} & {0.052${}\pm{}$0.049} &  {0.074${\,}\pm$} & {0.018${}\pm{}$0.010} &  {0.452${\,}\pm$} & {0.041${}\pm{}$0.054} \\
9.00 & 10.00 & 
 {0.086${\,}\pm$} & {0.020${}\pm{}$0.010} &  {0.0082${\,}\pm$} & {0.0053${}\pm{}$0.0012} &  {0.083${\,}\pm$} & {0.017${}\pm{}$0.010} \\
10.00 & 12.00 & 
 & {} &  {0.00095${\,}\pm$} & {0.00095${}\pm{}$0.00015} &  {0.0060${\,}\pm$} & {0.0028${}\pm{}$0.0008} \\
\hline
\hline
\end{tabular}
\label{table_xs_pt_be}
\end{table}

\begin{table}[h]
\caption{Invariant differential cross sections $\left( \DIFFXS \right)$ 
for direct-photon production in \pp\ collisions at 800 and
530~GeV/$c$, and \pip\ collisions at 515~GeV/$c$ as functions of $p_T$.}
\begin{tabular}{r@{ -- }l r@{ }l r@{ }l r@{ }l}
\hline
\hline
\multicolumn{2}{c}{$p_T$}   &\multicolumn{2}{c}{${\mit p}$${\mit p}$ at 800~GeV/$c$}   &\multicolumn{2}{c}{${\mit p}$${\mit p}$ at 530~GeV/$c$}   &\multicolumn{2}{c}{${\mit \pi^{-}}$${\mit p}$ at 515~GeV/$c$} \\
\multicolumn{2}{c}{ }   &\multicolumn{2}{c}{$-1.0\le\ycm\le 0.5$}   &\multicolumn{2}{c}{$-0.75\le\ycm\le 0.75$}   &\multicolumn{2}{c}{$-0.75\le\ycm\le 0.75$} \\
\multicolumn{2}{c}{(GeV/$c$)}    &\multicolumn{2}{c}{$[\rm{nb}$/(GeV/$c)^2]$}    &\multicolumn{2}{c}{$[\rm{nb}$/(GeV/$c)^2]$}    &\multicolumn{2}{c}{$[\rm{nb}$/(GeV/$c)^2]$}  \\
\hline
3.50 & 4.00 & 
 {1.88${\,}\pm$} & {0.30${}\pm{}$0.38} &  {1.16${\,}\pm$} & {0.19${}\pm{}$0.22} &  {1.72${\,}\pm$} & {0.42${}\pm{}$0.31} \\
4.00 & 4.50 & 
 {0.600${\,}\pm$} & {0.044${}\pm{}$0.10} &  {0.375${\,}\pm$} & {0.030${}\pm{}$0.062} &  {0.493${\,}\pm$} & {0.028${}\pm{}$0.076} \\
4.50 & 5.00 & 
 {0.2255${\,}\pm$} & {0.0094${}\pm{}$0.034} &  {0.1076${\,}\pm$} & {0.0045${}\pm{}$0.016} &  {0.165${\,}\pm$} & {0.011${}\pm{}$0.022} \\
\hline
\multicolumn{2}{c}{ }    &\multicolumn{2}{c}{$[\rm{pb}$/(GeV/$c)^2]$}    &\multicolumn{2}{c}{$[\rm{pb}$/(GeV/$c)^2]$}    &\multicolumn{2}{c}{$[\rm{pb}$/(GeV/$c)^2]$}  \\
\hline
5.00 & 5.50 & 
 {82.2${\,}\pm$} & {4.4${}\pm{}$11} &  {33.6${\,}\pm$} & {2.2${}\pm{}$4.4} &  {64.8${\,}\pm$} & {6.0${}\pm{}$8.1} \\
5.50 & 6.00 & 
 {35.0${\,}\pm$} & {2.4${}\pm{}$4.4} &  {11.4${\,}\pm$} & {1.1${}\pm{}$1.4} &  {25.5${\,}\pm$} & {3.5${}\pm{}$3.0} \\
6.00 & 7.00 & 
 {9.80${\,}\pm$} & {0.74${}\pm{}$1.2} &  {3.43${\,}\pm$} & {0.37${}\pm{}$0.42} &  {9.0${\,}\pm$} & {1.3${}\pm{}$1.0} \\
7.00 & 8.00 & 
 {1.73${\,}\pm$} & {0.28${}\pm{}$0.20} &  {0.42${\,}\pm$} & {0.11${}\pm{}$0.05} &  {2.81${\,}\pm$} & {0.62${}\pm{}$0.33} \\
8.00 & 10.00 & 
 {0.339${\,}\pm$} & {0.070${}\pm{}$0.039} &  {0.010${\,}\pm$} & {0.015${}\pm{}$0.001} &  {0.30${\,}\pm$} & {0.14${}\pm{}$0.04} \\
10.00 & 12.00 & 
 {0.017${\,}\pm$} & {0.014${}\pm{}$0.002} &  & {} &  & {} \\
\hline
\hline
\end{tabular}
\label{table_xs_pt_p}
\end{table}

\newpage

\begin{table}[h]
\caption{The invariant differential cross section 
$\left( \DIFFXS \right)$ per nucleon as a function of \ycm\ and $p_T$
for the inclusive reaction \pBe~$\rightarrow\gamma$X at 530~GeV/$c$.}
\begin{tabular}{r@{ -- }lr@{ }lr@{ }lr@{ }l}
\hline
\hline
\multicolumn{2}{c}{$p_T$ } &\multicolumn{2}{c}{$-0.75$ $\le\ycm\le$ $-0.50$}&\multicolumn{2}{c}{$-0.50$ $\le\ycm\le$ $-0.25$}&\multicolumn{2}{c}{$-0.25$ $\le\ycm\le$ $0.00$} \\
\multicolumn{2}{c}{(GeV/$c$)} &\multicolumn{2}{c}{$[\rm{pb}$/(GeV/$c)^2$]}&\multicolumn{2}{c}{$[\rm{pb}$/(GeV/$c)^2$]}&\multicolumn{2}{c}{$[\rm{pb}$/(GeV/$c)^2$]} \\
\hline
3.50 & 4.00 & 
 {1150${\,}\pm$} & {220${}\pm{}$210} &  {1470${\,}\pm$} & {210${}\pm{}$270} &  {1390${\,}\pm$} & {220${}\pm{}$260} \\
4.00 & 4.50 & 
 {339${\,}\pm$} & {14${}\pm{}$57} &  {381${\,}\pm$} & {12${}\pm{}$64} &  {365${\,}\pm$} & {11${}\pm{}$62} \\
4.50 & 5.00 & 
 {87.9${\,}\pm$} & {4.3${}\pm{}$13} &  {105.5${\,}\pm$} & {4.5${}\pm{}$16} &  {105.4${\,}\pm$} & {4.7${}\pm{}$16} \\
5.00 & 5.50 & 
 {22.5${\,}\pm$} & {1.9${}\pm{}$3.0} &  {28.2${\,}\pm$} & {2.0${}\pm{}$3.8} &  {36.4${\,}\pm$} & {2.3${}\pm{}$4.9} \\
5.50 & 6.50 & 
 {5.19${\,}\pm$} & {0.48${}\pm{}$0.65} &  {7.41${\,}\pm$} & {0.60${}\pm{}$0.92} &  {8.25${\,}\pm$} & {0.68${}\pm{}$1.0} \\
6.50 & 8.00 & 
 {0.61${\,}\pm$} & {0.11${}\pm{}$0.08} &  {0.97${\,}\pm$} & {0.14${}\pm{}$0.12} &  {1.20${\,}\pm$} & {0.16${}\pm{}$0.15} \\
8.00 & 10.00 & 
 {0.020${\,}\pm$} & {0.016${}\pm{}$0.003} &  {0.034${\,}\pm$} & {0.019${}\pm{}$0.005} &  {0.038${\,}\pm$} & {0.020${}\pm{}$0.005} \\
\hline
\multicolumn{2}{c}{  } &\multicolumn{2}{c}{$0.00$ $\le\ycm\le$ $0.25$}&\multicolumn{2}{c}{$0.25$ $\le\ycm\le$ $0.50$}&\multicolumn{2}{c}{$0.50$ $\le\ycm\le$ $0.75$} \\
\hline
3.50 & 4.00 & 
 {1510${\,}\pm$} & {190${}\pm{}$280} &  {1570${\,}\pm$} & {160${}\pm{}$290} &  {1120${\,}\pm$} & {260${}\pm{}$200} \\
4.00 & 4.50 & 
 {370.8${\,}\pm$} & {9.9${}\pm{}$63} &  {367.2${\,}\pm$} & {9.3${}\pm{}$62} &  {290${\,}\pm$} & {52${}\pm{}$49} \\
4.50 & 5.00 & 
 {107.2${\,}\pm$} & {4.3${}\pm{}$16} &  {111.4${\,}\pm$} & {4.2${}\pm{}$16} &  {87.1${\,}\pm$} & {4.0${}\pm{}$13} \\
5.00 & 5.50 & 
 {37.2${\,}\pm$} & {2.2${}\pm{}$5.0} &  {36.8${\,}\pm$} & {2.1${}\pm{}$4.9} &  {28.1${\,}\pm$} & {2.0${}\pm{}$3.8} \\
5.50 & 6.50 & 
 {8.95${\,}\pm$} & {0.65${}\pm{}$1.1} &  {8.19${\,}\pm$} & {0.63${}\pm{}$1.0} &  {7.42${\,}\pm$} & {0.61${}\pm{}$0.92} \\
6.50 & 8.00 & 
 {0.85${\,}\pm$} & {0.14${}\pm{}$0.11} &  {1.07${\,}\pm$} & {0.15${}\pm{}$0.13} &  {0.62${\,}\pm$} & {0.13${}\pm{}$0.08} \\
8.00 & 10.00 & 
 {0.055${\,}\pm$} & {0.025${}\pm{}$0.007} &  {0.040${\,}\pm$} & {0.021${}\pm{}$0.005} &  {0.038${\,}\pm$} & {0.022${}\pm{}$0.005} \\
\hline
\hline
\end{tabular}
\label{table_xs_pt_rap_530_be}
\end{table}

\begin{table}[h]
\caption{The invariant differential cross section 
$\left( \DIFFXS \right)$ per nucleon as a function of \ycm\ and $p_T$
for the inclusive reaction \pBe~$\rightarrow\gamma$X at 800~GeV/$c$.}
\begin{tabular}{r@{ -- }lr@{ }lr@{ }lr@{ }l}
\hline
\hline
\multicolumn{2}{c}{$p_T$ } &\multicolumn{2}{c}{$-1.0$ $\le\ycm\le$ $-0.75$}&\multicolumn{2}{c}{$-0.75$ $\le\ycm\le$ $-0.50$}&\multicolumn{2}{c}{$-0.50$ $\le\ycm\le$ $-0.25$} \\
\multicolumn{2}{c}{(GeV/$c$)} &\multicolumn{2}{c}{$[\rm{pb}$/(GeV/$c)^2$]}&\multicolumn{2}{c}{$[\rm{pb}$/(GeV/$c)^2$]}&\multicolumn{2}{c}{$[\rm{pb}$/(GeV/$c)^2$]} \\
\hline
3.50 & 4.00 & 
 {1200${\,}\pm$} & {320${}\pm{}$230} &  {2000${\,}\pm$} & {320${}\pm{}$380} &  {2020${\,}\pm$} & {320${}\pm{}$380} \\
4.00 & 4.50 & 
 {516${\,}\pm$} & {43${}\pm{}$90} &  {499${\,}\pm$} & {44${}\pm{}$87} &  {703${\,}\pm$} & {44${}\pm{}$120} \\
4.50 & 5.00 & 
 {129.3${\,}\pm$} & {7.4${}\pm{}$20} &  {157.7${\,}\pm$} & {8.2${}\pm{}$24} &  {218.8${\,}\pm$} & {8.9${}\pm{}$33} \\
5.00 & 5.50 & 
 {44.2${\,}\pm$} & {3.8${}\pm{}$6.0} &  {64.6${\,}\pm$} & {4.0${}\pm{}$8.8} &  {78.2${\,}\pm$} & {4.6${}\pm{}$11} \\
5.50 & 6.50 & 
 {10.31${\,}\pm$} & {0.93${}\pm{}$1.3} &  {18.5${\,}\pm$} & {1.3${}\pm{}$2.3} &  {23.3${\,}\pm$} & {1.4${}\pm{}$2.9} \\
6.50 & 8.00 & 
 {1.35${\,}\pm$} & {0.18${}\pm{}$0.16} &  {1.93${\,}\pm$} & {0.24${}\pm{}$0.23} &  {3.41${\,}\pm$} & {0.32${}\pm{}$0.40} \\
8.00 & 10.00 & 
 {0.170${\,}\pm$} & {0.045${}\pm{}$0.020} &  {0.169${\,}\pm$} & {0.046${}\pm{}$0.020} &  {0.325${\,}\pm$} & {0.070${}\pm{}$0.038} \\
\hline
\multicolumn{2}{c}{  } &\multicolumn{2}{c}{$-0.25$ $\le\ycm\le$ $0.00$}&\multicolumn{2}{c}{$0.00$ $\le\ycm\le$ $0.25$}&\multicolumn{2}{c}{$0.25$ $\le\ycm\le$ $0.50$} \\
\hline
3.50 & 4.00 & 
 {3710${\,}\pm$} & {410${}\pm{}$700} &  {2680${\,}\pm$} & {250${}\pm{}$510} &  {1870${\,}\pm$} & {250${}\pm{}$350} \\
4.00 & 4.50 & 
 {792${\,}\pm$} & {39${}\pm{}$140} &  {898${\,}\pm$} & {38${}\pm{}$160} &  {763${\,}\pm$} & {35${}\pm{}$130} \\
4.50 & 5.00 & 
 {234.7${\,}\pm$} & {8.5${}\pm{}$36} &  {297${\,}\pm$} & {12${}\pm{}$45} &  {250${\,}\pm$} & {12${}\pm{}$38} \\
5.00 & 5.50 & 
 {95.5${\,}\pm$} & {4.6${}\pm{}$13} &  {100.5${\,}\pm$} & {5.5${}\pm{}$14} &  {103.4${\,}\pm$} & {5.6${}\pm{}$14} \\
5.50 & 6.50 & 
 {27.2${\,}\pm$} & {1.4${}\pm{}$3.4} &  {30.7${\,}\pm$} & {1.6${}\pm{}$3.8} &  {25.1${\,}\pm$} & {1.6${}\pm{}$3.1} \\
6.50 & 8.00 & 
 {3.78${\,}\pm$} & {0.32${}\pm{}$0.44} &  {4.19${\,}\pm$} & {0.37${}\pm{}$0.49} &  {3.29${\,}\pm$} & {0.37${}\pm{}$0.38} \\
8.00 & 10.00 & 
 {0.149${\,}\pm$} & {0.054${}\pm{}$0.017} &  {0.285${\,}\pm$} & {0.076${}\pm{}$0.033} &  {0.412${\,}\pm$} & {0.100${}\pm{}$0.048} \\
\hline
\hline
\end{tabular}
\label{table_xs_pt_rap_800_be}
\end{table}

\newpage

\begin{table}[h]
\caption{The invariant differential cross section 
$\left( \DIFFXS \right)$ per nucleon as a function of \ycm\ and $p_T$
for the inclusive reaction \piBe~$\rightarrow\gamma$X at 515~GeV/$c$.}
\begin{tabular}{r@{ -- }lr@{ }lr@{ }lr@{ }l}
\hline
\hline
\multicolumn{2}{c}{$p_T$ } &\multicolumn{2}{c}{$-0.75$ $\le\ycm\le$ $-0.50$}&\multicolumn{2}{c}{$-0.50$ $\le\ycm\le$ $-0.25$}&\multicolumn{2}{c}{$-0.25$ $\le\ycm\le$ $0.00$} \\
\multicolumn{2}{c}{(GeV/$c$)} &\multicolumn{2}{c}{$[\rm{pb}$/(GeV/$c)^2$]}&\multicolumn{2}{c}{$[\rm{pb}$/(GeV/$c)^2$]}&\multicolumn{2}{c}{$[\rm{pb}$/(GeV/$c)^2$]} \\
\hline
3.50 & 4.00 & 
 {1250${\,}\pm$} & {220${}\pm{}$210} &  {1150${\,}\pm$} & {100${}\pm{}$190} &  {1312${\,}\pm$} & {88${}\pm{}$220} \\
4.00 & 4.50 & 
 {370${\,}\pm$} & {15${}\pm{}$58} &  {342${\,}\pm$} & {11${}\pm{}$53} &  {401${\,}\pm$} & {11${}\pm{}$63} \\
4.50 & 5.00 & 
 {80.4${\,}\pm$} & {5.0${}\pm{}$11} &  {124.3${\,}\pm$} & {4.7${}\pm{}$17} &  {137.4${\,}\pm$} & {5.1${}\pm{}$19} \\
5.00 & 5.50 & 
 {30.7${\,}\pm$} & {2.2${}\pm{}$3.9} &  {48.0${\,}\pm$} & {2.5${}\pm{}$6.0} &  {54.6${\,}\pm$} & {2.7${}\pm{}$6.8} \\
5.50 & 6.50 & 
 {8.63${\,}\pm$} & {0.64${}\pm{}$1.0} &  {12.79${\,}\pm$} & {0.79${}\pm{}$1.5} &  {18.51${\,}\pm$} & {0.93${}\pm{}$2.2} \\
6.50 & 8.00 & 
 {0.96${\,}\pm$} & {0.14${}\pm{}$0.11} &  {1.76${\,}\pm$} & {0.19${}\pm{}$0.20} &  {2.84${\,}\pm$} & {0.24${}\pm{}$0.33} \\
8.00 & 10.00 & 
 {0.060${\,}\pm$} & {0.029${}\pm{}$0.007} &  {0.159${\,}\pm$} & {0.040${}\pm{}$0.019} &  {0.294${\,}\pm$} & {0.056${}\pm{}$0.035} \\
\hline
\multicolumn{2}{c}{  } &\multicolumn{2}{c}{$0.00$ $\le\ycm\le$ $0.25$}&\multicolumn{2}{c}{$0.25$ $\le\ycm\le$ $0.50$}&\multicolumn{2}{c}{$0.50$ $\le\ycm\le$ $0.75$} \\
\hline
3.50 & 4.00 & 
 {1670${\,}\pm$} & {120${}\pm{}$280} &  {1490${\,}\pm$} & {100${}\pm{}$250} &  {1240${\,}\pm$} & {67${}\pm{}$210} \\
4.00 & 4.50 & 
 {436${\,}\pm$} & {11${}\pm{}$68} &  {444${\,}\pm$} & {11${}\pm{}$69} &  {387${\,}\pm$} & {11${}\pm{}$61} \\
4.50 & 5.00 & 
 {162.7${\,}\pm$} & {5.1${}\pm{}$22} &  {175.2${\,}\pm$} & {5.4${}\pm{}$24} &  {165.5${\,}\pm$} & {5.4${}\pm{}$23} \\
5.00 & 5.50 & 
 {64.1${\,}\pm$} & {2.7${}\pm{}$8.1} &  {73.2${\,}\pm$} & {2.9${}\pm{}$9.2} &  {69.1${\,}\pm$} & {3.0${}\pm{}$8.7} \\
5.50 & 6.50 & 
 {21.73${\,}\pm$} & {0.95${}\pm{}$2.6} &  {23.30${\,}\pm$} & {0.99${}\pm{}$2.8} &  {21.9${\,}\pm$} & {1.0${}\pm{}$2.6} \\
6.50 & 8.00 & 
 {3.22${\,}\pm$} & {0.25${}\pm{}$0.37} &  {4.09${\,}\pm$} & {0.29${}\pm{}$0.47} &  {3.10${\,}\pm$} & {0.28${}\pm{}$0.36} \\
8.00 & 10.00 & 
 {0.424${\,}\pm$} & {0.065${}\pm{}$0.051} &  {0.367${\,}\pm$} & {0.068${}\pm{}$0.044} &  {0.237${\,}\pm$} & {0.053${}\pm{}$0.028} \\
\hline
\hline
\end{tabular}
\label{table_xs_pt_rap_515_be}
\end{table}

\begin{table}[h]
\caption{The invariant differential cross section 
$\left( \DIFFXS \right)$ as a function of \ycm\ and $p_T$
for the inclusive reaction \pp~$\rightarrow\gamma$X at 530~GeV/$c$.}
\begin{tabular}{r@{ -- }lr@{ }lr@{ }lr@{ }l}
\hline
\hline
\multicolumn{2}{c}{$p_T$ } &\multicolumn{2}{c}{$-0.75$ $\le\ycm\le$ $-0.50$}&\multicolumn{2}{c}{$-0.50$ $\le\ycm\le$ $-0.25$}&\multicolumn{2}{c}{$-0.25$ $\le\ycm\le$ $0.00$} \\
\multicolumn{2}{c}{(GeV/$c$)} &\multicolumn{2}{c}{$[\rm{pb}$/(GeV/$c)^2$]}&\multicolumn{2}{c}{$[\rm{pb}$/(GeV/$c)^2$]}&\multicolumn{2}{c}{$[\rm{pb}$/(GeV/$c)^2$]} \\
\hline
3.50 & 4.00 & 
 {2210${\,}\pm$} & {570${}\pm{}$400} &  {1180${\,}\pm$} & {480${}\pm{}$220} &  {850${\,}\pm$} & {420${}\pm{}$160} \\
4.00 & 4.50 & 
 {356${\,}\pm$} & {32${}\pm{}$60} &  {363${\,}\pm$} & {28${}\pm{}$61} &  {411${\,}\pm$} & {25${}\pm{}$69} \\
4.50 & 5.00 & 
 {93${\,}\pm$} & {12${}\pm{}$14} &  {112${\,}\pm$} & {11${}\pm{}$16} &  {128${\,}\pm$} & {11${}\pm{}$19} \\
5.00 & 5.50 & 
 {24.8${\,}\pm$} & {5.1${}\pm{}$3.3} &  {35.8${\,}\pm$} & {5.0${}\pm{}$4.8} &  {38.0${\,}\pm$} & {5.6${}\pm{}$5.1} \\
5.50 & 6.50 & 
 {5.6${\,}\pm$} & {1.2${}\pm{}$0.7} &  {9.3${\,}\pm$} & {1.6${}\pm{}$1.2} &  {6.9${\,}\pm$} & {1.6${}\pm{}$0.9} \\
6.50 & 8.00 & 
 {0.25${\,}\pm$} & {0.20${}\pm{}$0.03} &  {0.84${\,}\pm$} & {0.32${}\pm{}$0.10} &  {0.87${\,}\pm$} & {0.34${}\pm{}$0.11} \\
\hline
\multicolumn{2}{c}{  } &\multicolumn{2}{c}{$0.00$ $\le\ycm\le$ $0.25$}&\multicolumn{2}{c}{$0.25$ $\le\ycm\le$ $0.50$}&\multicolumn{2}{c}{$0.50$ $\le\ycm\le$ $0.75$} \\
\hline
3.50 & 4.00 & 
 {1110${\,}\pm$} & {390${}\pm{}$200} &  {800${\,}\pm$} & {240${}\pm{}$150} &  {560${\,}\pm$} & {520${}\pm{}$100} \\
4.00 & 4.50 & 
 {315${\,}\pm$} & {23${}\pm{}$53} &  {327${\,}\pm$} & {22${}\pm{}$55} &  {390${\,}\pm$} & {160${}\pm{}$70} \\
4.50 & 5.00 & 
 {110${\,}\pm$} & {11${}\pm{}$16} &  {103.7${\,}\pm$} & {9.6${}\pm{}$15} &  {72.9${\,}\pm$} & {9.2${}\pm{}$11} \\
5.00 & 5.50 & 
 {36.0${\,}\pm$} & {5.3${}\pm{}$4.8} &  {33.0${\,}\pm$} & {5.1${}\pm{}$4.4} &  {24.2${\,}\pm$} & {4.7${}\pm{}$3.2} \\
5.50 & 6.50 & 
 {10.4${\,}\pm$} & {1.7${}\pm{}$1.3} &  {10.0${\,}\pm$} & {1.6${}\pm{}$1.2} &  {4.0${\,}\pm$} & {1.3${}\pm{}$0.5} \\
6.50 & 8.00 & 
 {1.01${\,}\pm$} & {0.35${}\pm{}$0.12} &  {1.65${\,}\pm$} & {0.41${}\pm{}$0.20} &  {0.60${\,}\pm$} & {0.27${}\pm{}$0.07} \\
\hline
\hline
\end{tabular}
\label{table_xs_pt_rap_530_p}
\end{table}

\newpage

\begin{table}[h]
\caption{The invariant differential cross section 
$\left( \DIFFXS \right)$ as a function of \ycm\ and $p_T$
for the inclusive reaction \pp~$\rightarrow\gamma$X at 800~GeV/$c$.}
\begin{tabular}{r@{ -- }lr@{ }lr@{ }lr@{ }l}
\hline
\hline
\multicolumn{2}{c}{$p_T$ } &\multicolumn{2}{c}{$-1.0$ $\le\ycm\le$ $-0.75$}&\multicolumn{2}{c}{$-0.75$ $\le\ycm\le$ $-0.50$}&\multicolumn{2}{c}{$-0.50$ $\le\ycm\le$ $-0.25$} \\
\multicolumn{2}{c}{(GeV/$c$)} &\multicolumn{2}{c}{$[\rm{pb}$/(GeV/$c)^2$]}&\multicolumn{2}{c}{$[\rm{pb}$/(GeV/$c)^2$]}&\multicolumn{2}{c}{$[\rm{pb}$/(GeV/$c)^2$]} \\
\hline
3.50 & 4.00 & 
 {1990${\,}\pm$} & {870${}\pm{}$380} &  {1320${\,}\pm$} & {790${}\pm{}$250} &  {990${\,}\pm$} & {780${}\pm{}$190} \\
4.00 & 4.50 & 
 {483${\,}\pm$} & {72${}\pm{}$84} &  &  &  {744${\,}\pm$} & {83${}\pm{}$130} \\
4.50 & 5.00 & 
 {162${\,}\pm$} & {20${}\pm{}$25} &  {211${\,}\pm$} & {20${}\pm{}$32} &  {219${\,}\pm$} & {21${}\pm{}$33} \\
5.00 & 5.50 & 
 {44.0${\,}\pm$} & {7.6${}\pm{}$6.0} &  {78.9${\,}\pm$} & {9.6${}\pm{}$11} &  {84${\,}\pm$} & {11${}\pm{}$11} \\
5.50 & 6.50 & 
 {12.6${\,}\pm$} & {2.3${}\pm{}$1.6} &  {20.8${\,}\pm$} & {3.0${}\pm{}$2.6} &  {25.0${\,}\pm$} & {3.3${}\pm{}$3.1} \\
6.50 & 8.00 & 
 {1.69${\,}\pm$} & {0.49${}\pm{}$0.20} &  {2.20${\,}\pm$} & {0.64${}\pm{}$0.26} &  {3.02${\,}\pm$} & {0.79${}\pm{}$0.35} \\
8.00 & 10.00 & 
 {0.18${\,}\pm$} & {0.11${}\pm{}$0.02} &  {0.45${\,}\pm$} & {0.20${}\pm{}$0.05} &  {0.44${\,}\pm$} & {0.19${}\pm{}$0.05} \\
\hline
\multicolumn{2}{c}{  } &\multicolumn{2}{c}{$-0.25$ $\le\ycm\le$ $0.00$}&\multicolumn{2}{c}{$0.00$ $\le\ycm\le$ $0.25$}&\multicolumn{2}{c}{$0.25$ $\le\ycm\le$ $0.50$} \\
\hline
3.50 & 4.00 & 
 {2290${\,}\pm$} & {810${}\pm{}$430} &  {2510${\,}\pm$} & {570${}\pm{}$470} &  {2190${\,}\pm$} & {550${}\pm{}$410} \\
4.00 & 4.50 & 
 {801${\,}\pm$} & {52${}\pm{}$140} &  {810${\,}\pm$} & {100${}\pm{}$140} &  {621${\,}\pm$} & {79${}\pm{}$110} \\
4.50 & 5.00 & 
 {208${\,}\pm$} & {19${}\pm{}$31} &  {330${\,}\pm$} & {27${}\pm{}$50} &  {222${\,}\pm$} & {30${}\pm{}$34} \\
5.00 & 5.50 & 
 {85${\,}\pm$} & {10${}\pm{}$12} &  {111${\,}\pm$} & {13${}\pm{}$15} &  {90${\,}\pm$} & {13${}\pm{}$12} \\
5.50 & 6.50 & 
 {30.4${\,}\pm$} & {3.5${}\pm{}$3.8} &  {31.5${\,}\pm$} & {3.8${}\pm{}$3.9} &  {23.7${\,}\pm$} & {3.8${}\pm{}$2.9} \\
6.50 & 8.00 & 
 {3.74${\,}\pm$} & {0.79${}\pm{}$0.44} &  {5.28${\,}\pm$} & {0.96${}\pm{}$0.62} &  {4.18${\,}\pm$} & {0.94${}\pm{}$0.49} \\
8.00 & 10.00 & 
 {0.25${\,}\pm$} & {0.16${}\pm{}$0.03} &  &  &  {0.77${\,}\pm$} & {0.23${}\pm{}$0.09} \\
\hline
\hline
\end{tabular}
\label{table_xs_pt_rap_800_p}
\end{table}

\begin{table}[h]
\caption{The invariant differential cross section 
$\left( \DIFFXS \right)$ as a function of \ycm\ and $p_T$
for the inclusive reaction \pip~$\rightarrow\gamma$X at 515~GeV/$c$.}
\begin{tabular}{r@{ -- }lr@{ }lr@{ }lr@{ }l}
\hline
\hline
\multicolumn{2}{c}{$p_T$ } &\multicolumn{2}{c}{$-0.75$ $\le\ycm\le$ $-0.50$}&\multicolumn{2}{c}{$-0.50$ $\le\ycm\le$ $-0.25$}&\multicolumn{2}{c}{$-0.25$ $\le\ycm\le$ $0.00$} \\
\multicolumn{2}{c}{(GeV/$c$)} &\multicolumn{2}{c}{$[\rm{pb}$/(GeV/$c)^2$]}&\multicolumn{2}{c}{$[\rm{pb}$/(GeV/$c)^2$]}&\multicolumn{2}{c}{$[\rm{pb}$/(GeV/$c)^2$]} \\
\hline
4.00 & 4.50 & 
 {477${\,}\pm$} & {76${}\pm{}$74} &  {454${\,}\pm$} & {79${}\pm{}$71} &  {566${\,}\pm$} & {71${}\pm{}$88} \\
4.50 & 5.00 & 
 {156${\,}\pm$} & {26${}\pm{}$21} &  {158${\,}\pm$} & {27${}\pm{}$22} &  {189${\,}\pm$} & {29${}\pm{}$26} \\
5.00 & 5.50 & 
 {43${\,}\pm$} & {12${}\pm{}$5.0} &  {52${\,}\pm$} & {14${}\pm{}$6.0} &  {73${\,}\pm$} & {16${}\pm{}$9.0} \\
5.50 & 6.50 & 
 {6.5${\,}\pm$} & {3.5${}\pm{}$0.8} &  {14.2${\,}\pm$} & {4.6${}\pm{}$1.7} &  {20.5${\,}\pm$} & {5.0${}\pm{}$2.4} \\
6.50 & 8.00 & 
 {2.1${\,}\pm$} & {1.0${}\pm{}$0.2} &  {2.9${\,}\pm$} & {1.2${}\pm{}$0.3} &  {4.0${\,}\pm$} & {1.5${}\pm{}$0.5} \\
\hline
\multicolumn{2}{c}{  } &\multicolumn{2}{c}{$0.00$ $\le\ycm\le$ $0.25$}&\multicolumn{2}{c}{$0.25$ $\le\ycm\le$ $0.50$}&\multicolumn{2}{c}{$0.50$ $\le\ycm\le$ $0.75$} \\
\hline
4.00 & 4.50 & 
 {514${\,}\pm$} & {64${}\pm{}$80} &  {547${\,}\pm$} & {57${}\pm{}$85} &  {403${\,}\pm$} & {66${}\pm{}$63} \\
4.50 & 5.00 & 
 {137${\,}\pm$} & {26${}\pm{}$19} &  {208${\,}\pm$} & {29${}\pm{}$29} &  {145${\,}\pm$} & {29${}\pm{}$20} \\
5.00 & 5.50 & 
 {65${\,}\pm$} & {14${}\pm{}$8.0} &  {84${\,}\pm$} & {16${}\pm{}$11} &  {72${\,}\pm$} & {16${}\pm{}$9.0} \\
5.50 & 6.50 & 
 {25.9${\,}\pm$} & {5.3${}\pm{}$3.1} &  {25.0${\,}\pm$} & {5.5${}\pm{}$3.0} &  {19.8${\,}\pm$} & {5.5${}\pm{}$2.3} \\
6.50 & 8.00 & 
 {4.7${\,}\pm$} & {1.6${}\pm{}$0.5} &  {4.7${\,}\pm$} & {1.6${}\pm{}$0.5} &  {4.9${\,}\pm$} & {1.9${}\pm{}$0.6} \\
\hline
\hline
\end{tabular}
\label{table_xs_pt_rap_515_p}
\end{table}

\end{document}